\documentclass[superscriptaddress,onecolumn,secnumarabic,
amssymb,amsmath,nobibnotes,aps,prd,showkeys,showpacs,nofootinbib]{revtex4}

\usepackage[latin1]{inputenc}
\usepackage{graphicx}
\usepackage[english]{babel}

\usepackage{amsmath}
\usepackage{amssymb}
\usepackage{amsfonts}
\usepackage{colordvi}
\usepackage{psfrag}
\usepackage{color}

\allowdisplaybreaks[4]

\begin{document}

\newcommand\cL{\mathcal{L}}
\newcommand\be{\begin{equation}}
\newcommand\ee{\end{equation}}
\newcommand\bea{\begin{eqnarray}}
\newcommand\eea{\end{eqnarray}}
\newcommand\beq{\begin{eqnarray}}
\newcommand\eeq{\end{eqnarray}}
\newcommand\tr{{\rm tr}\, }
\newcommand\nn{\nonumber \\}
\newcommand\e{{\rm e}}

\newcommand\bef{\begin{figure}}
\newcommand\eef{\end{figure}}
\newcommand{\ans}{ansatz }
\newcommand{\eeqn}{\end{eqnarray}}
\newcommand{\bd}{\begin{displaymath}}
\newcommand{\ed}{\end{displaymath}}
\newcommand{\mat}[4]{\left(\begin{array}{cc}{#1}&{#2}\\{#3}&{#4}
\end{array}\right)}
\newcommand{\matr}[9]{\left(\begin{array}{ccc}{#1}&{#2}&{#3}\\
{#4}&{#5}&{#6}\\{#7}&{#8}&{#9}\end{array}\right)}
\newcommand{\matrr}[6]{\left(\begin{array}{cc}{#1}&{#2}\\
{#3}&{#4}\\{#5}&{#6}\end{array}\right)}
\newcommand{\cvb}[3]{#1^{#2}_{#3}}
\newcommand\lsim{\raise0.3ex\hbox{$\;<$\kern-0.75em\raise-1.1ex
e\hbox{$\sim\;$}}}
\newcommand\gsim{\raise0.3ex\hbox{$\;>$\kern-0.75em\raise-1.1ex
\hbox{$\sim\;$}}}
\def\abs#1{\left| #1\right|}
\newcommand\simlt{\mathrel{\lower2.5pt\vbox{\lineskip=0pt\baselineskip=0pt
           \hbox{$<$}\hbox{$\sim$}}}}
\newcommand\simgt{\mathrel{\lower2.5pt\vbox{\lineskip=0pt\baselineskip=0pt
           \hbox{$>$}\hbox{$\sim$}}}}
\newcommand\unity{{\hbox{1\kern-.8mm l}}}
\newcommand{\eps}{\varepsilon}
\newcommand\ep{\epsilon}
\newcommand\ga{\gamma}
\newcommand\Ga{\Gamma}
\newcommand\om{\omega}
\newcommand\omp{{\omega^\prime}}
\newcommand\Om{\Omega}
\newcommand\la{\lambda}
\newcommand\La{\Lambda}
\newcommand\al{\alpha}
\newcommand{\ov}{\overline}
\renewcommand{\to}{\rightarrow}
\renewcommand{\vec}[1]{\mathbf{#1}}
\newcommand{\vect}[1]{\mbox{\boldmath$#1$}}
\newcommand\tm{{\widetilde{m}}}
\newcommand\mcirc{{\stackrel{o}{m}}}
\newcommand{\Dm}{\Delta m}
\newcommand{\dm}{\varepsilon}
\newcommand{\tanb}{\tan\beta}
\newcommand{\nbar}{\tilde{n}}
\newcommand\PM[1]{\begin{pmatrix}#1\end{pmatrix}}
\newcommand{\up}{\uparrow}
\newcommand{\down}{\downarrow}
\newcommand\omE{\omega_{\rm Ter}}
%

\newcommand{\Dsusy}{{susy \hspace{-9.4pt} \slash}\;}
\newcommand{\DCP}{{CP \hspace{-7.4pt} \slash}\;}
\newcommand{\mc}{\mathcal}
\newcommand{\gr}{\mathbf}
\renewcommand{\to}{\rightarrow}
\newcommand{\gtc}{\mathfrak}
\newcommand{\wh}{\widehat}
\newcommand{\br}{\langle}
\newcommand{\kt}{\rangle}


\def\lsim{\mathrel{\mathop  {\hbox{\lower0.5ex\hbox{$\sim$}
\kern-0.8em\lower-0.7ex\hbox{$<$}}}}}
\def\gsim{\mathrel{\mathop  {\hbox{\lower0.5ex\hbox{$\sim$}
\kern-0.8em\lower-0.7ex\hbox{$>$}}}}}

\newcommand\de{\partial}
\newcommand\brf{{\mathbf f}}
\newcommand\bbf{\bar{\bf f}}
\newcommand\bF{{\bf F}}
\newcommand\bbF{\bar{\bf F}}
\newcommand\bA{{\mathbf A}}
\newcommand\bB{{\mathbf B}}
\newcommand\bG{{\mathbf G}}
\newcommand\bI{{\mathbf I}}
\newcommand\bM{{\mathbf M}}
\newcommand\bY{{\mathbf Y}}
\newcommand\bX{{\mathbf X}}
\newcommand\bS{{\mathbf S}}
\newcommand\bb{{\mathbf b}}
\newcommand\bh{{\mathbf h}}
\newcommand\bg{{\mathbf g}}
\newcommand\bla{{\mathbf \la}}
\newcommand\bmu{\mathbf m }
\newcommand\by{{\mathbf y}}
\newcommand\bsig{\mbox{\boldmath $\sigma$} }
\newcommand\bunity{{\mathbf 1}}
\newcommand\cA{\mathcal{A}}
\newcommand\cB{\mathcal{B}}
\newcommand\cC{\mathcal{C}}
\newcommand\cD{\mathcal{D}}
\newcommand\cF{\mathcal{F}}
\newcommand\cG{\mathcal{G}}
\newcommand\cH{\mathcal{H}}
\newcommand\cI{\mathcal{I}}
\newcommand\cN{\mathcal{N}}
\newcommand\cM{\mathcal{M}}
\newcommand\cO{\mathcal{O}}
\newcommand\cR{\mathcal{R}}
\newcommand\cS{\mathcal{S}}
\newcommand\cT{\mathcal{T}}
\newcommand\eV{\mathrm{eV}}

\title{Evaporation and Antievaporation instabilities}

\author{Andrea Addazi}
\author {Antonino Marciano}
\affiliation{Center for Field Theory and Particle Physics \& Department of Physics, Fudan University, 200433 Shanghai, China}


\begin{abstract}
\noindent 
We review (anti)evaporation phenomena within the context of quantum gravity and extended theories of gravity. The (anti)evaporation effect is an instability of the black hole horizon discovered in many different scenarios: quantum dilaton-gravity, $f(R)$-gravity, $f(T)$-gravity, string inspired black holes and brane-world cosmology. Evaporating and antievaporating black holes seem to have completely different thermodynamical features compared to standard semiclassical black holes. The purpose of this review is to provide an introduction to conceptual and technical aspects of (anti)evaporation effects, while discussing problems that are still open.

\end{abstract}


\maketitle

\section{Introduction}
\noindent 
The long-standing idea to extend the standard model of Einsteinian gravity, General Relativity (GR), is strongly motivated by several open issues in cosmology and quantum gravity. Despite several known successful applications of GR to astrophysics and cosmology, its UV completion and some cosmological and astrophysical instantiations, including the inflationary paradigm and the comprehension of the nature of dark energy and dark matter, still remain puzzling. The most popular extension of GR remains $f(R)$-gravity, including $(R+\zeta R^{2})$ Starobinsky's model for inflation \cite{Nojiri:2006ri,Nojiri:2010wj,Clifton:2011jh,Capozziello:2011et,Nojiri:2017ncd}. This theory can be conformally mapped onto scalar-tensor theories or dilaton-gravity theories \cite{Nojiri:2006ri,Nojiri:2010wj,Clifton:2011jh,Capozziello:2011et}, in regular unambiguous space-time backgrounds. There are many alternatives that have been hitherto suggested, such as $f(T)$-gravity \cite{Cai:2015emx}, Mimetic Gravity \cite{Chamseddine:2014vna,Sebastiani:2016ras,Ijjas:2016pad,Rabochaya:2015haa,Nojiri:2014zqa}, string-inspired black holes and brane-world cosmologies \cite{ArkaniHamed:1999hk,Randall:1999ee,Randall:1999vf,Dvali:2000hr} --- see \cite{Maartens:2010ar,Brax:2003fv} for reviews on brane-world
cosmological scenarios --- just to mention few of them. 

We review aspects of instabilities of a class of black hole solutions, which appear universally in these aforementioned classes of extended theories of gravity, and are dubbed (anti)evaporation instabilities. (Anti)evaporation phenomena consist in the exponentially (growing) decreasing radius of the black hole horizon. These were first discovered by {\it Bousso} and {\it Hawking} within the context of quantum dilaton-gravity --- see e.g. Ref~\cite{Bousso:1997wi} --- and then elaborated in Refs.~\cite{Nojiri:1998ph,Nojiri:1998ue,Elizalde:1999dw}. {\it Nojiri} and {\it Odintsov} rediscovered the same effect in $f(R)$-gravity at the classical level in Ref.~\cite{Nojiri:2013su,Nojiri:2014jqa} --- see also Ref.\cite{Sebastiani:2013fsa} for technical improvements. The two phenomena were further studied in several other contexts, such as Gauss-Bonnet gravity \cite{Sebastiani:2013fsa}, $f(T)$-gravity \cite{Houndjo:2013qna}, Mimetic Gravity \cite{Oikonomou:2016fxb,Matsumoto:2015wja,Astashenok:2015haa,Oikonomou:2015lgy}, Bigravity \cite{Katsuragawa:2014hda}, string-inspired black hole solutions \cite{Addazi:2016hip}, brane-world cosmology \cite{Chakraborty:2014xla,Chakraborty:2015bja,Chakraborty:2016ydo,Addazi:2017puj} and Bardeen De Sitter black holes \cite{Singh:2017qur}. In all these theories, the two metric solutions, which turned out to be unstable, are Nariai, a degenerate Schwarzschild-De Sitter black hole, and extremal Reissner-N\"ordstrom solutions, in which two horizons coincide. In Ref.~\cite{Addazi:2016prb}, through the analysis of the Raychaduri equation, describing the dynamics of BH closed trapped Cauchy's surfaces --- similar technics were used in General Relativity in Refs.~\cite{Ellis:2013oka,Firouzjaee:2014zfa,Firouzjaee:2015bqa} --- it was argued that classical (anti)evaporation instabilities switch off the emission of Bekenstein-Hawking radiation \cite{Bekenstein,Hawking:1974sw,Hawking,Hawking2}. 
Very recently, the (anti-)evaporation was also discussed in relation with energy conditions in extended theories 
of gravity \cite{Addazi:2017vti}. 

Among all the possible scenarios, it is worth to mention that there are many 
realistic extensions of general relativity which are compatible with cosmological 
and astrophysical limits and which predict anti-evaporation phenomena. 
Certainly the minimal and more appealing scenarios 
seem the ones provided by $f(R)$-gravity models. 
For example, among all possible $f(R)$-gravity extensions, 
some simple models already proposed in literature
-- and well compatible with cosmological 
constraints --
such as the Hu-Sawicki model, exponential $f(R)$-gravity 
and higher derivative polynomial extensions beyond Starobinsky's gravity 
universally exhibit the (anti-)evaporation phenomena
(see Ref.\cite{Sebastiani:2013fsa} for a detailed discussion on these aspects).

The plan of the paper is the following. 
In Sec. 2 we briefly introduce the concept of evaporation and antievaporation instabilities. 
In Sec. 3 we review the (anti)evaporation in quantum dilaton-gravity;
in Sec. 4, we review the classical (anti-)evaporation in $f(R)$-gravity; 
in Sec. 5, we review (anti-)evaporation in $f(T)$-gravity; 
in Sec. 6, we review the classical (anti-)evaporation phenomena in context of string-inspired black holes; 
in Sec. 7, we review either Hawking?s radiation in (anti-)evaporating black holes;
in Sec. 8 we review classical (anti-)evaporation of FRW brane-worlds sourced by (anti-)evaporating instabilities of the higher dimensional black hole in the bulk. 
In Sec. 9, we show our conclusions and remarks. 

\section{What is (Anti)evaporation?}

The evaporation and the antievaporation  
are related to a dynamical decreasing and increasing of the black hole 
horizon radius in time. These instabilities may be provoked by several different dynamical origins. 
The possible sources of them can be classified in two kinds: i) quantum anomalies; ii) classical instabilities 
sourced by extensions of General Relativity. In the next sections, we will review many possible models with 
(anti)evaporation instabilities, lying in (i,ii) classes.

\section{(Anti)evaporation in quantum dilaton-gravity} \label{aeqdg}

\noindent
In this section we review studies and results obtained on antievaporation within the context of quantum dilaton-gravity  \cite{Bousso:1997wi,Nojiri:1998ph,Nojiri:1998ue,Elizalde:1999dw}.

We start considering the four-dimensional action of $N$ scalars fields coupled to gravity, which are included in the theory in order to allow the description of black hole radiation. The action then reads 
\be \label{SSS}
S=\frac{1}{16\pi G_N}\int d^{4}x \sqrt{-g^{(4)}}\left[R^{(4)}-2\lambda-\frac{1}{2}\sum_{1}^{N}(\nabla^{(4)}\Phi_{i})^{2}\right]\, ,
\ee
where $G_N$ is the Newton constant, $\Phi_{i}$ are N-scalar fields, and $g^{(4)}$, $R^{(4)}$ and $\nabla^{(4)}\Phi_{i}$ are respectively the four-dimensional metric determinant, the covariant derivative with respect to the four-dimensional metric and the Ricci scalar. \\

We consider the spherically symmetric background {\it ansatz}
\be \label{pertNaraiQD}
ds^{2}=e^{2\rho(x,t)}(dx^{2}-dt^{2})+e^{-2\phi(x,t)}d\Omega^{2}\,,
\ee
in which $\phi(x,t),\rho(x,t)$ are functions of space-time coordinates and $d\Omega^{2}$ is the two-dimensional angular line-element. In the background (\ref{pertNaraiQD}), the integration of the angular modes can be performed. The 4D action reduces to a two-dimensional one, which reads 
\be \label{twodim}
S=\frac{1}{16\pi }\int d^{2}x\sqrt{-g}e^{-2\phi}[R+2(\nabla \phi)^{2}+2e^{2\phi}-2\lambda-\sum_{i=1}^{N}(\nabla \Phi_{i})^{2} ]\,.
\ee

 It was shown in \cite{CF} that the amount of black hole radiation at infinity is proportional to the trace anomaly. The trace of the energy-momentum tensor is classically vanishing, but if we consider the quantum nature of fields, a non-vanishing expectation value of the trace can be recovered on curved background. The inclusion of the trace anomaly in the dynamics of the system under scrutiny accommodates the analysis of the back reaction of the evaporation on the geometry. This is equivalent to take into account the one-loop effective action of the matter field.

Following the same strategy as in \cite{Bousso:1997cg}, two-dimensional conformal scalar fields with exponential dilation coupling yield the the trace anomaly 
\be
T=\frac{1}{24 \pi} \left[ R - 6 (\nabla \phi)^2 - 2 \partial^{2} \phi \right]\,.
\ee
The trace anomaly can be obtained from using the zeta
function 
approach and general proprieties of the trace anomaly  \cite{Bousso:1997cg}.

Equivalently, from \eqref{twodim} the 
the scale-dependent part of the one-loop effective action for dilaton coupled scalars
reads 
\be \label{dilcou}
S_{1}=-\frac{1}{48\pi}\int d^{2}x\sqrt{-g}[\frac{1}{2 } R \frac{1}{ \partial^{2} } R]
-6(\nabla \phi)^{2}\frac{1}{\partial^{2}}R-2\phi R ]\, .
\ee

As shown in \cite{Hayward:1994dw}, the action \eqref{dilcou} can be recast as local by introducing an auxiliary scalar field $A$ that mimics the trace anomaly term. 
In other words, the trace anomaly derived from the effective action \eqref{dilcou}. 

As shown in \cite{Hayward:1994dw}, 
 the action \eqref{dilcou} can be rewritten in the following form:
\be \label{obt}
S=\frac{1}{16\pi}\int d^{2}x\sqrt{-g}\left[\left( e^{-2\phi}+\frac{\kappa}{2}(A+w\phi)\right)R-\frac{\kappa}{4}(\nabla A)^{2}
+2+2e^{-2\phi}(\nabla \phi)^{2}-2e^{-2\phi}\lambda\right]\,,
\ee
where $\kappa=2N/3$ and $w$ is a numerical factor. In the large N-limit, the quantum fluctuations of the metric are dominated by the quantum fluctuations of the N scalars, thus $\kappa>\!\!>1$.
Such a formal rewriting is possible in the framework of the scalar auxiliar field method \cite{Hayward:1994dw}.

We can now derive the effective dynamics of the system. Variations of the effective action with respect to $\rho, \phi$ and $A$ lead to 
\be 
\label{llelel}
-\left(1-\frac{w\kappa}{4}e^{2\phi}\right)\partial^{2}\phi+2(\partial \phi)^{2}+\frac{\kappa}{4}e^{2\phi}\partial^{2}A
+e^{2\rho+2\phi}(\lambda e^{-2\phi}-1)=0\,,
\ee
\be 
\label{eomsm}
\left(1-\frac{w\kappa}{4}e^{2\phi}\right)\partial^{2}\rho-\partial^{2}\phi+(\partial \phi)^{2}+\lambda e^{2\rho}=0\,,
\ee
\be
\label{eomsm2}
\partial^{2}A-2\partial^{2}\rho=0\, , 
\ee
with additional two constraints to be considered  {\it i.e.}

\be
\label{lelell}
\left( 1-\frac{w\kappa}{4}e^{2\phi}\right)(\delta^{2}\phi-2\delta \phi \delta \rho)-(\delta \phi)^{2}=\frac{\kappa}{8}e^{2\phi}[(\delta A)^{2}+2\delta^{2}A-4\delta A\delta \rho]\, ,
\ee
\be \label{secondcontst}
\left( 1-\frac{w\kappa}{4}e^{2\phi}\right)(\dot{\phi}'-\dot{\rho}\phi'-\rho'\dot{\phi})-\rho'\dot{\phi}=\frac{\kappa}{8}e^{2\phi}\left[\dot{A}A'+2\dot{A}'-2(\dot{\rho}A'+\rho'\dot{A})   \right]\, ,
\ee
having used the conventions 
$$\partial A \partial B=-\dot{A}\dot{B}+f'g',\,\,\,\partial^{2}g=-\ddot{A}+B''\, ,$$
$$\delta A \delta B=\dot{A}\dot{B}+A'B',\,\,\,\delta^{2}A=\ddot{A}+A''\, .$$

From Eq.\ref{eomsm2}, one obtains 
\be
\label{ZZZ}
A=2\rho+\eta\, ,
\ee
with $\eta$ any harmonic function of $x$ and $t$. Relation \eqref{ZZZ} eliminates the dependence by $A$ in the other EoMs. 

In such a formalism, we can study perturbations around the Nariai solution (See Appendix A). The Nariai solution, which corresponds to Eq.(\ref{pertNaraiQD}) with $e^{-\phi}={\rm const}$, is a solution of the dilaton-gravity theory that reads 
\be \label{Nariai}
e^{2\rho}=\frac{1}{\Lambda_{1}}\frac{1}{\cos^{2}t},\,\,\,\,\,\,e^{2\phi}=\Lambda_{2}\, , 
\ee
where 
\be \label{Lambdau}
\frac{1}{\Lambda_{1}}=\frac{1}{8\Lambda}\left[4-(w+2)b+\sqrt{16-8(w-2)+(w+2)^{2}b^{2}}    \right]\, ,
\ee
\be \label{Lambdad}
\Lambda_{2}=\frac{1}{2w\kappa}\left[ 4+(w+2)b-\sqrt{16-8(w-2)b+(w+2)^{2}b^{2}}\right]\, . 
\ee
In these latter we have defined $b=\kappa \Lambda$, and assumed $b<\!\!<1$ for $\kappa>\!\!>1$. \\

We may perturb this solution around the Nariai background, and obtain
\be \label{variation}
e^{2\phi}=\Lambda_{2}[1+2\epsilon \sigma(t)\cos x]\, ,
\ee
where $\epsilon<\!\!<1$. We might also perturb $e^{2\rho}$, but contributions that would arise from $e^{2\rho}$ would not enter the equation of motion for $\sigma$ at the first order of the $\epsilon$-expansion. \\

Let us now consider the condition for a black hole horizon $(\nabla \phi)^{2}=0$. Substituting in this latter relation Eq.(\ref{variation}), we obtain a simple system of differential equations, {\it i.e.} 
\be\label{phidd}
\frac{\partial \phi}{\partial t}=\epsilon \dot{\sigma}\cos x\, ,\,\,\,\, \frac{\partial \phi}{\partial x}=-\epsilon \sigma \sin x\, .
\ee
At the first order in the $\epsilon$-expansion, the black hole radius casts 
\be \label{rBBB}
r_{b}(t)^{-2}=e^{2\phi}=\lambda_{2}[1+2\epsilon \delta(t)]\, ,
\ee
\be \label{horiocn}
\delta \equiv \cos x_{b}=\sigma\left(1+\frac{\dot{\sigma}^{2}}{\sigma^{2}}\right)^{-1/2}\, .
\ee
Consequently the black hole horizon is controlled by the equation of motion for $\sigma$
\be \label{eom}
\frac{\ddot{\sigma}}{\sigma}=\frac{a}{\cos^{2}t}-1\, ,
\ee
where 
\be \label{aa}
a=\frac{2\sqrt{16-8(w-2)b+(w+2)^{2}b^{2}}}{4-wb}\, . 
\ee
The classical limit is obtained when we send $\kappa \rightarrow 0$. 
In this limit the equation of motion is exactly solvable, and reduces to 
\be \label{EoMss}
\dot{\sigma}=\sigma \tan t\, ,
\ee
which yields the solution 
\be \label{sigmaz}
\sigma(t)=\frac{\sigma_{0}}{\cos t}
\ee
for the initial condition $\dot{\sigma}_{0}(t=0)=0$. This leads to a perturbation $\delta(t)=\sigma_{0}={\rm const}$, which ensures that the Nariai solution is static at classical level. Nonetheless, at quantum level, for $\kappa>0$, we obtain an approximated solution for the perturbations 
\be \label{deltat}
\delta(t)=\sigma_{0}\left[1-\frac{1}{2}(a-1)(a-2)t^{2}+O(t^{4})\right],\,\,\,\, \sigma_{0}>0\,,\,\,\,\dot{\sigma}(t=0)=0\, . 
\ee
As a remarkable consequence, the black hole size increases, {\it i.e.} the maximal Schwarzschild-de Sitter black hole has an anti-evaporation instability.

\section{(Anti)-evaporation in $f(R)$-gravity}
\noindent 
In this section, we review some basic aspects of evaporation and antievaporation in $f(R)$-gravity \cite{Nojiri:2013su,Nojiri:2014jqa,Sebastiani:2013fsa}, taking into account the Nariai metric and extremal Reissner-Nodrstr\"om black holes. Let us first recall the theoretical framework.

In $f(R)$-gravity, the action reads \cite{Nojiri:2006ri,Nojiri:2010wj,Clifton:2011jh,Capozziello:2011et,Nojiri:2017ncd}
\be \label{action}
I=\frac{1}{16\pi}\int d^{4}x\sqrt{-g}f(R)+S_{m}\,,
\ee
written in units $G_{N}=c=1$. Varying the action Eq.(\ref{action}) with respect to the metric tensor, we obtain the equation of motions (EoMs) of the theory 
\be \label{EoM}
\frac{1}{2}g_{\mu\nu}f(R)-f'(R)R_{\mu\nu}+\nabla_{\mu}\nabla_{\nu}f'(R)=-8\pi T^{m}_{\mu\nu}\,.
\ee
Whenever the matter content is vanishing, namely $T^{m}_{\mu\nu}=0$, and the Ricci tensor constant, {\it i.e.} $R_{\mu\nu}\sim g_{\mu\nu}$, the EoM is reduced to a more manageable form, {\it i.e.}
\be \label{EoM2}
f(R)-\frac{1}{2}Rf'(R)=0\,.
\ee

\subsection{The case of the Nariai black hole in $f(R)$-gravity}
\noindent
The Nariai space-time is a solution of Eq.\eqref{EoM2}. It can be recast --- for details see Appendix A --- as
\be \label{Narai}
ds^{2}=\frac{1}{\Lambda^{2}}\left(\frac{1}{\cos h^{2}x}(dx^{2}-dt^{2})+d\Omega^{2}\right)\,,
\ee
where $\Lambda$ has one mass dimension, and again $d\Omega^{2}$ denotes the solid angle on a 2-sphere, {\it i.e.} $d\Omega^{2}=d\theta^{2}+\sin^{2}\theta d\phi^{2}$, with $\theta\in[0,\pi)$ and $\phi\in[0,2\pi)$. Notice also that the Ricci scalar of the Nariai space-time is $R_{0}=4\Lambda^{2}={\rm const}$.

The Nariai metric can be obtained from the 
 more general expression 
\be \label{pertNarai}
ds^{2}=e^{2\rho(x,t)}(dx^{2}-dt^{2})+e^{-2\phi(x,t)}d\Omega^{2}\,,
\ee
where $\phi(x,t),\rho(x,t)$ are functions of space-time coordinates 
governed by the following EoMs:
\be \label{FF}
0=-\frac{e^{2\rho}}{2}f(R)-(-\ddot{\rho}+2\ddot{\phi}+\rho''-2\dot{\phi}^{2}-2\rho'\phi'-2\dot{\rho}\dot{\phi})f'(R)+\frac{\partial^{2}f'}{\partial t^{2}}-\dot{\rho}\frac{\partial f'}{\partial t}-\rho'\frac{\partial f'}{\partial x}
\ee
$$+e^{2\phi}\left\{-\frac{\partial}{\partial t}\left(e^{-2\phi}\frac{\partial f'}{\partial t} \right)+\frac{\partial}{\partial x}\left(e^{-2\phi}\frac{\partial f'}{\partial x} \right)  \right\}\, ,$$
\be \label{FF1}
0=\frac{e^{2\rho}}{2}f-(\ddot{\rho}+2\phi''-\rho''-2\phi'^{2}-2\rho'\phi'-2\dot{\rho}\dot{\phi})f'+\frac{\partial^{2}f'}{\partial x^{2}}
-\dot{\rho}\frac{\partial f'}{\partial t}-\rho'\frac{\partial f'}{\partial x}
\ee
$$-e^{2\phi}\left\{-\frac{\partial}{\partial t}\left(e^{-2\phi}\frac{\partial f'}{\partial t} \right)+\frac{\partial}{\partial x}\left( 
e^{-2\phi} \frac{\partial f'}{\partial x}\right)  \right\}\, ,$$
\be \label{FF2}
0=-(2\dot{\phi}'-2\phi'\dot{\phi}-2\rho'\dot{\phi}-2\dot{\rho}\phi')f'+\frac{\partial^{2}f'}{\partial t \partial x}
-\dot{\rho}\frac{\partial f'}{\partial x}-\rho'\frac{\partial f'}{\partial t}\, , 
\ee
\be \label{FF3}
0=\frac{e^{-2\phi}}{2}f-e^{-2(\rho+\phi)}(-\ddot{\phi}+\phi''-2\phi'^{2}+2\dot{\phi}^{2})f'-f'
+e^{-2(\rho+\phi)}\left(\dot{\phi}\frac{\partial f'}{\partial t}-\phi' \frac{\partial f'}{\partial x}\right)
\ee
$$-e^{-2\rho}\left\{ -\frac{\partial}{\partial t}\left( e^{-2\phi}\frac{\partial f'}{\partial t}\right)+\frac{\partial}{\partial}\left( e^{-2\phi}\frac{\partial f'}{\partial x}\right)   \right\}\, .$$

From EoMs in the metric (\ref{pertNarai}), one can study the evolution of the perturbations around the Nariai background: 
\be \label{rhooo}
\rho=-{\rm ln}\,(\Lambda \cosh x)+\delta \rho\, ,
\ee
\be 
\label{phiii}\phi={\rm ln}\,  \Lambda+\delta \phi\, .
\ee
Substituting these expressions into EoMs, one obtains a set of four equations in $\delta \rho,\delta \phi$, namely   
\be \label{perf1}
0=\frac{-f'(R_{0})+2\Lambda^{2}f''(R_{0})}{2\Lambda^{2}\cosh^{2}x}\delta R-\frac{f(R_{0})}{\Lambda^{2}\cosh^{2}x}\delta \rho
\ee
$$-f'(R_{0})(-\delta \ddot{\rho}+2\delta \ddot{\phi}+\delta \rho''+2\tanh x \delta \phi')+\tanh x f''(R_{0})\delta R'+f''(R_{0})\delta R'' \, ,$$
\be \label{perf2}
0=-\frac{-f'(R_{0})+2\Lambda^{2}f''(R_{0})}{2\Lambda^{2}\cosh^{2}x}\delta R+\frac{f(R_{0})}{\Lambda^{2}\cosh^{2}x}\delta \rho
\ee
$$-f'(R_{0})(\delta \ddot{\rho}+2\delta \phi''-\delta \rho''+2\tanh x \delta \phi')
+f''(R_{0})\delta \ddot{R}+\tanh x f''(R_{0})\delta R'\, ,$$
\be \label{perf3}
0=-2(\delta \dot{\phi}'+\tanh x \delta \dot{\phi})+\frac{f''(R_{0})}{f'(R_{0})}(\delta \dot{R}'+\tanh x \delta \dot{R})\, , 
\ee
\be \label{perf4}
0=-\frac{-f'(R_{0})+2\Lambda^{2}f''(R_{0})}{2\Lambda^{2}}\delta R-\frac{f(R_{0})}{\Lambda^{2}}\delta \phi-\cosh^{2}xf'(R_{0})
(-\delta \ddot{\phi}+\delta \phi'')-\cosh^{2}x f''(R_{0})(-\delta \ddot{R}+\delta R'')\,,
\ee
where 
\be \label{RRRR}
\delta R=4\Lambda^{2}(-\delta \rho+\delta \phi)+\Lambda^{2}\cosh^{2}x(2\delta \ddot{\rho}-2\delta\rho''-4\delta \ddot{\phi}+\delta \phi'')\,.
\ee

The third equation can be integrated, leading to 
\be \label{thirdin}
-2\delta \phi+\frac{f''(R_{0})}{f'(R_{0})}\delta R=c_{x}(x)+\frac{c_{t}(t)}{\cosh x}\, ,
\ee
where $c_{x}(x),c_{t}(t)$ are arbitrary integration functions of 
$x,t$ respectively. 
From a linear combination of the first, second and fourth equations, 
one can obtain the 
equations
\be \label{oobbe}
0=\frac{-f'(R_{0})+2\Lambda^{2}f''(R_{0})}{2\Lambda^{2}\cosh^{2}x}\delta R-f'(R_{0})\partial^{2} \left(\delta \rho-\delta \phi-\frac{f''(R_{0})}{f'(R_{0})}\delta R \right)\, ,
\ee
\be \label{oeen}
0=\frac{2\Lambda^{2}}{\cosh^{2}x}\delta \phi+\partial^{2} \left( \delta \rho+\frac{f''(R_{0})}{2f'(R_{0})}\delta R\right)\, .
\ee
Eqs.~(\ref{oobbe}), (\ref{oeen}), once combined with Eq.~(\ref{thirdin}), allow to find the differential equation in $\phi$ 
\be \label{uniq}
0=\frac{1}{\alpha \cosh^{2}x}\left(2(2\alpha-1)\delta \phi+(\alpha-1)\left(c_{x}(x)+\frac{c_{t}(t)}{\cosh x} \right)\right)
+\partial^{2}\left(3\delta \phi+c_{x}(x)+\frac{c_{t}(t)}{\cosh x} \right)\, ,
\ee
where 
\be \label{alphaaa}
\alpha\equiv \frac{2\Lambda^{2}f''(R_{0})}{f'(R_{0})}\,.
\ee
We emphasize that Eq.~(\ref{uniq}) can have unstable modes in specific subregions of the parameter space. \\

Since in homogeneous and isotropic backgrounds $\delta \phi(t,x)\equiv \phi(t)$, Eq.~(\ref{uniq}) reduces to 
\be \label{deltaphi}
\frac{d^{2}\delta \phi}{dt^{2}}+\tanh t \frac{d \delta \phi}{\delta t}-m^{2}\delta \phi=0\, ,
\ee
where the effective mass of the mode is expressed by
\be\label{mass}
m^{2}=\frac{2(2\alpha-1)}{3\alpha}\, ,
\ee
having assumed the initial conditions $c_{x}=c_{t}=0$ in Eq.~(\ref{uniq}). Such an equation has tachyon-like modes for $m^{2}>0$ ($\alpha<0$ and $\alpha > 1/2$) and for $1+4m^{2}\geq 0$ ($\alpha<0$ and $\alpha > 8/19$).

The horizon is located in correspondence of the condition 
\be \label{cond}
\nabla \delta \phi \cdot \nabla \delta \phi=0\,,
\ee
which specifies the requirement that the gradient of the two-sphere size is equal to zero. This means that for a black hole located in $x_{0}$, the radius is 
$$r_{0}(t)^{-2}=e^{2\phi(t,x_{0})}\, .$$
Consequently, either an increase or a decrease of $\phi$ correspond to a dynamical displacement of the horizon. 

\subsection{Extremal Reissner-Nordstr\"om black holes}

In this section we will review evaporation and antievaporation of the extremal Reissner-N\"orstrom (RN) black holes in $f(R)$-gravity \cite{Nojiri:2014jqa}. 

The extremal RN solution is recovered in the limit in which the two possible RN radii coincide. The extremal RN-black hole metric can be then recast as --- see Appendix B for further details --- 
\be \label{Mass}
ds^{2}=\frac{r_{0}^{2}}{\left( 1-\frac{r_{0}^{2}R_{0}}{2}\right)\cosh^{2}x}(d\tau^{2}-dx^{2})+r_{0}^{2}d\Omega^{2}\,.
\ee
This expression shares several similarities with the aforementioned Nariai metric. Indeed the extremal RN solution also can reshuffled as
\be \label{asf}
ds^{2}=\frac{e^{2\rho(x,\tau)}}{\Lambda^{2}}(d\tau^{2}-dx^{2})+\frac{e^{-2\phi(x,\tau)}}{\Lambda'^{2}}d\Omega^{2}\, .
\ee
The form of $\rho(x,\tau)$ finally induces the explicit formula
\be \label{LambdaL}
ds^{2}=\frac{1}{\Lambda^{2}\cosh^{2}x}(d\tau^{2}-dx^{2})+\frac{e^{-2\phi}}{\Lambda'^{2}}d\Omega^{2}\, ,
\ee
\be \label{LLL}
\Lambda=\frac{\sqrt{1-\frac{r_{0}^{2}R_{0}}{2}}}{r_{0}},\,\,\,\,\Lambda'=\frac{1}{r_{0}}\, .
\ee
Using then the same {\it ansatz} on the metric we deployed while tackling the Nariai metric, the EoM, written in components $(\tau,\tau),(x,x),(\tau,x)$ and $(\theta,\theta)$ $((\psi,\psi))$, cast
\be \label{Eom1}
0=\frac{e^{2\rho}}{2\Lambda^{2}}f(R)-\left( -\ddot{\rho}+2\ddot{\phi}+\rho''-2\dot{\phi}^{2}-2\rho'\phi'-2\dot{\rho}\dot{\phi}\right)f'(R)+\frac{\partial^{2}f'(R)}{\partial \tau^{2}}-\dot{\rho}\frac{\partial f'}{\partial \tau}-\rho'\frac{\partial f'}{\partial x}
\ee
$$+e^{2\phi}\left\{-\frac{\partial}{\partial \tau}\left(e^{-2\phi}\frac{\partial f'}{\partial \tau} \right)+\frac{\partial}{\partial x}\left(e^{-2\phi}\frac{\partial F'}{\partial x} \right)   \right\}\,,$$
\be \label{Eom2}
0=\frac{e^{2\rho}}{2\Lambda^{2}}f(R)-(\ddot{\rho}+2\phi''-\rho''-2\phi'^{2}-2\rho'\phi'-2\dot{\rho}\dot{\phi})f'+\frac{\partial^{2}f'}{\partial x^{2}}-\dot{\rho}\frac{\partial f'}{\partial \tau}-\rho' \frac{\partial f'}{\partial x}
\ee
$$-e^{2\phi}\left\{ -\frac{\partial}{\partial \tau}\left(e^{-2\phi}\frac{\partial f'}{\partial \tau} \right)+\frac{\partial}{\partial x}
\left(e^{-2\phi}\frac{\partial f'}{\partial x} \right)\right\}\, ,$$
\be \label{Eom3}
0=-(2\dot{\phi}'-2\phi' \dot{\phi}-2\rho' \dot{\phi}-2\dot{\rho} \phi')f'+\frac{\partial^{2}}{\partial \tau \partial x}-\dot{\rho}\frac{\partial f'}{\partial x}-\rho' \frac{\partial f'}{\partial \tau}\, ,
\ee
\be \label{Eom4}
0=-\frac{e^{-2\phi}}{2\Lambda'^{2}}f-\frac{\Lambda^{2}}{\Lambda'^{2}}e^{-2(\rho+\phi)}\left(-\ddot{\phi}+\phi''-2\phi'^{2}+2\dot{\phi}^{2} \right)f'+f'
\ee
$$+\frac{\Lambda^{2}}{\Lambda'^{2}}e^{-2(\rho+\phi)}\left(\dot{\phi}\frac{\partial f'}{\partial \tau}-\phi'\frac{\partial f'}{\partial x} \right)-\frac{\Lambda^{2}}{\Lambda'^{2}}e^{-2\rho}\left\{-\frac{\partial }{\partial \tau}\left( e^{-2\phi}\frac{\partial f'}{\partial \tau}\right)
+\frac{\partial}{\partial x}\left(e^{-2\phi}\frac{\partial f'}{\partial x} \right)  \right\}\, . $$

Perturbations with respect to the extremal RN background can be considered following the same strategy as in the previous sections. We then add a generic perturbation to the expressions
\be \label{perrt}
\rho=-\log \cosh x+\delta \rho,\,\,\, \phi=\delta \phi\, 
\ee
and then recover
\be \label{e1}
0=f''(R_{0})\left\{ -\frac{1}{\cosh^{2} x}\delta R+\tanh x \,\delta R'+\delta R''\right\}\, ,
\ee
\be \label{e2}
0=f''(R_{0})\left\{\frac{1}{\cosh^{2} x}\delta R+\tanh x \, \delta R'+\delta \ddot{R}\right\}\, ,
\ee
\be \label{e3}
0=f''(R_{0})\left\{\delta \dot{R}'+\tanh x \delta \dot{R}\right\}\, ,
\ee
\be \label{e4}
0=f''(R_{0})\left\{\delta R-\cosh^{2} x(-\delta \ddot{R}+\delta R'')\right\}\, ,
\ee
where 
\be \label{dR}
 \delta R=-4\Lambda^{2}\delta \rho+4\Lambda'^{2}\delta \phi-\Lambda^{2}\cosh^{2}x\{2(\delta \ddot{\rho}-2\delta \rho'')-4(\delta
 \ddot{\phi}-\delta \phi'')\}\,.
\ee

To study the instabilities of the system, we can adopt the parametrization 
\be \label{deltarho}
\delta \phi=\phi_{0}\cosh \omega \tau \cosh^{\beta}x,\qquad \,\,\,\,\delta\rho=\rho_{0}\cosh \omega \tau \cosh^{\beta}x\,,
\ee
where $\rho_{0},\phi_{0},\omega,\beta$ are constant parameters. 
Using the definition of  the horizon
$g^{\mu\nu}\nabla_{\mu}\phi \nabla_{\nu}\phi=0$,
we then end up recovering the solutions 
\be \label{dpho}
\delta \phi\equiv \delta \phi_{H}=\phi_{0}\cosh^{2}\beta t\, ,
\ee
\be \label{rhh}
r_{H}=\frac{1}{\Lambda}e^{-\delta \phi_{H}}=\frac{e^{-\phi_{0}\cosh^{2}\beta \tau}}{\Lambda}\, .
\ee
What is remarkable in this case is that the instabilities seem to {\it independent} by the particular kind of $f(R)$-gravity under scrutiny.

\section{(Anti)-evaporation in $f(T)$-gravity}
\noindent 
In this section we move to the discussion of the evaporation and antievaporation phenomena within the context of $f(T)$-gravity \cite{Houndjo:2013qna}. Once again, we start reviewing the theoretical framework of these models. \\ 

In $f(T)$-gravity, the action reads 
\be \label{actionft}
{I}=\frac{1}{16\pi}\int d^{4}x\sqrt{-g}f(T)+S_{m},
\ee
in which again we use units $G_N=c=1$.  We then introduce internal indices in the description of the gravitational field, and represent the gravitational degrees of freedom in terms of a frame field that constitutes the tetrad matrix. The line element then recasts
\be \label{ds}
ds^{2}=g_{\mu\nu}dx^{\mu}dx^{\nu}=\eta_{ij}\theta^{i}\theta^{j}\,,
\ee
\be \label{dx}
dx^{\mu}=e_{i}^{\mu}\theta^{i},\,\,\,\,\,\,\,\theta^{i}=e^{i}_{\mu}dx^{\mu}\,,
\ee
where $e_{i}^{\mu}e^{i}_{\nu}=\delta_{\nu}^{\mu}$, $\eta_{ij}={\rm diag}(-1,1,1,1)$,
$\sqrt{-g}=e={\rm det}[e_{\mu}^{i}]$. \\

The Weitzenb\"ock connection deployed in the construction of the $f(T)$ theory is purely torsional. Its relation to the torsion tensor can be straightforwardly determined to be
\be \label{T}
T_{\mu\nu}^{\alpha}=\Gamma_{\nu\mu}^{\alpha}-\Gamma_{\mu\nu}^{\alpha}=e^{\alpha}_{j}(\partial_{\mu}e_{\nu}^{i}-\partial_{\nu}e_{\mu}^{i})\, .
\ee
The Euler-Lagrange equations of the theory are then recovered by variation of the action with respect to the tetrad field $e_{\mu}^{i}$, namely 
\be \label{EOM}
S_{\mu}^{\nu\rho}\partial_{\rho}T \frac{d^{2}f}{dT^{2}}+e^{-1}e_{\mu}^{i}\partial_{\rho}[e S_{\alpha}^{\nu\rho}e_{i}^{\alpha}+T^{\alpha}_{\mu\sigma}S^{\nu\sigma}_{\alpha}]\frac{df}{dT}+\frac{1}{2}\delta_{\mu}^{\nu}f=4\pi T^{(m)}_{\mu\nu}\,.
\ee
In \eqref{EOM} $T^{(m)}_{\mu\nu}$ denotes the energy-momentum tensor, while $S_{\mu}^{\nu\rho}$ is expressed by the relation 
\be \label{S}
S_{\alpha}^{\mu\nu}=\frac{1}{2}(\delta_{\alpha}^{\mu}T^{\nu\beta}_{\beta}-\delta_{\beta}^{\mu}T^{\nu\beta}_{\alpha}+K_{\alpha}^{\mu\nu})\,,
\ee
$K_{\alpha}^{\mu\nu}$ standing for the co-torsion. Finally, the scalar torsion reads 
\be \label{S}
T=T_{\mu\nu}^{\alpha}S_{\alpha}^{\mu\nu}\,. 
\ee
General relativity with a cosmological constant can be recovered in the limit $\frac{d^{2}f}{dT^{2}}\rightarrow 0$, {\it i.e.} $f(T)=a + bT$.

\subsection{The case of Nariai Black hole in diagonal tetrads gauge}
\noindent 
For the $f(T)$ theory, the Nariai space-time acquires the form
\be \label{Narai}
ds^{2}=\frac{1}{\Lambda}\left[-\frac{1}{\cos^{2}\tau}(dx^{2}-d\tau^{2})+d\Omega^{2}\right]\,,
\ee
where $\Lambda$ is the cosmological constant, once again $d\Omega^{2}$ stands for the 
solid angle on a 2-sphere $d\Omega^{2}=d\theta^{2}+\sin^{2}\theta d\psi^{2}$, and $0<\tau<\pi/2$, $0<t<\infty$, with the mutual relation $\cosh t=1/\cos \tau$. Notice that also in this case the Ricci scalar of the Nariai space-time is constant, since $R=4\Lambda$. \\

The Nariai space-time is a solution of Eq.~(\ref{EOM}) in the diagonal tetrad {\it ansatz} 
\be \label{pertNarai}
ds^{2}=e^{2\rho(x,t)}(-dx^{2}+d\tau^{2})+e^{-2\phi(x,t)}d\Omega^{2}\, ,
\ee
\be \label{pertNarai2}
e_{\mu}^{a}=[ e^{\rho},e^{\rho},e^{-\phi},e^{-\phi}\sin \theta]\, .
\ee

The dynamical aspects of the Nariai solutions can be studied resorting to the methods of perturbation theory. We can consider arbitrary variations to the functions 
 \be \label{pertNarai3}
\rho=-ln[\sqrt{\Lambda}\cos \tau]+\delta \rho(\tau,x)\,,
\ee
\be \label{pertNarai4}
\phi={\rm ln}\sqrt{\Lambda}+\delta \phi(\tau,x)\,,
\ee
and then find the relation 
\be \label{pertNarai5}
\delta T=-2\Lambda \sin(2\tau)\delta \dot{\phi}\,.
\ee
Inserting Eqs.~(\ref{pertNarai4}) and (\ref{pertNarai5}) in Eq.~(\ref{EOM}), we may recover 
\be \label{pertNarai6}
\delta \phi(x,\tau)=k_{1}\sin(x-\bar{x}) \sec \tau+k_{2}\,,
\ee
where $\bar{x}$ is the fixed initial condition and $k_{1,2,}$ are two integration constants. 

Consider now that the horizon is defined through the condition 
\be \label{pertNarai7}
\left(\frac{\partial \delta \phi}{\partial \tau}\right)^{2}=\left(\frac{\partial \delta \phi}{\partial x}\right)^{2} \, .
\ee
From this, we obtain 
\be \label{pertNaraia8}
x_{h}=\bar{x}-\tau+m\pi-\frac{\pi}{2}\, ,
\ee
where $m=0,1,...$, and correspondingly we recover  
\be \label{pertNaraib}
\delta\phi(\tau,x_{h})=k_{1}(-1)^{n+1}+k_{2}\, ,
\ee
\be \label{pertNaraic}
r_{h}(\tau)^{-2}=1+ \delta\phi(\tau,x_{h})\, .
\ee
We can interpret this result saying that the black hole radius is fixed, {\it i.e.} no evaporation or antievaporation instabilities occur. 

It is worth to note that the diagonal tetrads choice should be handled carefully 
in the case of spherically symmetric solutions. 
This issue was extensively studied in Refs. \cite{Gonzalez:2011dr,Capozziello:2012zj,Iorio:2012cm,Krssak:2015oua}.
In Ref. \cite{Krssak:2015oua}, it was shown that the rigorous way to 
implement the tetrads choice consists in taking into account also the connection. 
These arguments highly motivate to relax the diagonal tetrads choice,
as discussed in the following section.

\subsection{Classical Evaporation and Antievaporation in non-diagonal tetrads}
\noindent
We can now generalize the previous result, considering a non-diagonal tetrad of the form 
\be \label{e00}
e_{0}^{0}=e^{\rho},\,\,\,e_{3}^{3}=e_{0}^{1,2,3}=e_{1,2,3}^{0}=0\, ,
\ee
\be \label{e11}
e_{1}^{1}=\cos \psi \sin \theta \,e^{\rho}\, ,\,\,\,
e_{1}^{2}=\cos \psi \cos \theta \,e^{-\phi}\, ,\,\,\,
e_{1}^{3}=-\sin \psi \sin \theta \,e^{-\phi}\, ,
\ee
\be \label{e21}
e_{2}^{1}=\sin \psi \sin \theta \, e^{\rho}\, ,\,\,\,
e_{3}^{1}=\cos \theta \, e^{\rho}\, ,\,\,\,
e_{2}^{2}=\sin \psi \cos \theta \,e^{-\phi}\, , 
\ee
\be \label{e32}
e_{3}^{2}=\cos \psi \sin \theta \,e^{-\phi},\,\,\,
e_{2}^{3}=-\sin \theta \,e^{-\phi}	\, .
\ee

Under this {\it ansatz} we obtain 
 \be \label{phi}
\delta \phi=A\sec \tau \cos(x-\bar{x})+B(\tan \tau)^{3/2}e^{\frac{1+2\cos^{2}\tau}{4\cos^{4}\tau}}\, , 
\ee
where $A,B$  are integration constants. This entails for the horizon the expression 
 \be \label{xh}
x_{h}=\bar{x}-\tau+{\rm arcsin}\left(\frac{\cos^{2}\tau}{A}\frac{d}{d\tau}\varphi(\tau) \right) \, , 
\ee
where 
 \be \label{varphi}
\varphi(\tau)=B(\tan \tau)^{3/2}e^{\frac{1+2\cos^{2}\tau}{4\cos^{4}\tau}}\, .
\ee
Notice that Eq.~(\ref{phi}) has a divergence in $\tau \rightarrow \pi/2$ --- this is the extreme time-like angle excluded from the range of the Nariai solution. Depending on the integration constants, Eq.~(\ref{xh}) represents a solution either increasing or decreasing in time. The first class of instabilities corresponds to the classical antievaporation, while the second class to the classical evaporation.

\section{(Anti)-evaporation in string-inspired black holes}

We discuss dyonic black hole solutions
in the case of $f(R)$-gravity 
coupled with a dilaton and 
two gauge bosons. 
The study of such a model is highly motivated from string theory.
Our Black Hole solutions are extensions 
of the one firstly studied by 
Kallosh, Linde, Ort\'in, Peet and Van Proyen
(KLOPV) in Ref.\cite{Kallosh:1992ii}.
We will show that extreme solutions are unstable.
In particular, these solutions 
have
Bousso-Hawking-Nojiri-Odintsov
(anti)evaporation instabilities.

As is known, 
the low energy limit of a dimensionally reduced superstring theory dimensionally reduced to $d=4$
is $\mathcal{N}=4$ supergravity. 
There are two versions:
$SO(4)$ and $SU(4)$. 
The first one is invariant
under a (rigid) $SU(4)\times SU(1,1)$ symmetry. 
Black hole solutions of the reduced sector $U(1)^{2}$ 
were studied  by Kallosh, Linde, Ort\'in,
Peet and Van Proeyen (KLOPV)
in 
Ref. \cite{Kallosh:1992ii}.
In particular, they
consider 
$U(1)^{2}$ charged dilaton black holes.
These solutions are
Reissner-Nordstr\"om-like black holes,
or more precisely of dyonic black holes. 
In particular, the dilaton field is the real part 
of an initial complex scalar, while the 
imaginary part is an axion pseudoscalar field. 
They assumed the axion stabilized to a constant VEV. 
The effective bosonic action 
corresponds to the Einstein-Hilbert one
coupled with a dilaton field and 
two $U(1)$ fields.
Extreme limits of dyonic solutions are shown to saturate $\mathcal{N}=4$
supersymmetry in $d=4$. 
On the other hand, the presence of non-perturbative stringy effects 
could modify the effective action in the low energy limit. 
For instance, higher derivative terms may be generated by Euclidean D-brane or worldsheet instantons. 
In particular, the Einstein-Hilbert sector coupled to the dilaton and $U(1)$-fields can be 
extended from $R$ to an analytic function $f(R)$   (See Ref.\cite{Bianchi:2009ij} for a review on this subject)
\footnote{See Refs.\cite{Addazi:2015goa,Addazi:2015yna,Addazi:2016mtn,Addazi:2016xuh}
for recent investigations of E-brane instantons in particle physics. }. 

KLOPP solutions are particularly important in string theory. 
For instance the famous 
derivation of the Hawking BH entropy
from BPS microstates shown by 
 Strominger and Vafa 
is based on five dimensional KLOPP solutions \cite{Strominger:1996sh}.
The Vafa-Strominger result has inspired the 
so called fuzzball proposal,
which has the ambition to solve the BH information 
paradox \cite{Mathur:2009hf}. 

 It is worth 
to mention that the existence of modes' correlations 
inside the Hawking radiation was discussed in Ref.\cite{Zhang:2013sza}.
 On the other hand, the unitarity time evolution 
of quantum black hole formation and evaporation processes in the framework of the 
Bohr-like approach was studied in Ref.\cite{Corda:2013gva}.

In this paper, we will study black hole solutions 
in string inspired $f(R)$-gravity, coupled with a dilaton field and two gauge bosons
\footnote{It is conceivable that analysis of branes in higher dimensional $f(R)$-gravity 
(See Refs.\cite{Chakraborty:2014xla,Chakraborty:2015bja,Chakraborty:2016ydo}) 
may be 
connected to these issues.}. 
We assume that the asymptotic space-time is 
Minkowski's one. 
Let us clarify that we will not consider a $f(R)$-{\it supergravity} coupled
to gauge bosons and dilatons. 
In fact, it was recently shown that 
the only $f(R)$-supergravity which 
is not plagued by ghosts and tachyons is 
Starobinsky's supergravity
\cite{Ketov:2013dfa,Ferrara:2013pla}. 
Nevertheless, one can consider the case in which
higher derivative terms are generated 
by exotic instantons or fluxes after a spontaneous 
supersymmetry breaking mechanism. 
In this sense, our model, which 
has a stable vacuum; and it is 
not plagued by ghosts and tachyons, 
is inspired by string theory. 
Clearly, to calculate instantonic corrections 
from a realistic stringy model is, at the moment,  impossible.
We believe that this highly motivates our 
effective field theory analysis, 
in which coefficients inside the $f(R)$-functional
parametrize our ignorance about the string theory vacua. 
We will show that extreme dyonic solutions 
have Bousso-Hawking-Nojiri-Odintsov (BHNO)  
(anti)evaporation instabilities.
In particular, Nojiri and Odintsov have discovered
(anti)evaporation instabilities in 
Reissner-Nordstr\"om black holes 
in $f(R)$-gravity \cite{Nojiri:2014jqa}. 
 { \it A posteriori}, our result is understood 
as a generalization of Nojiri-Odintsov calculations 
in Ref. \cite{Nojiri:2014jqa}. 
On the other hand, the peculiar thermodynamical proprieties of antievaporating solutions 
were discussed in our recent paper 
 \cite{Addazi:2016prb}.

Let us consider the case of 
a $f(R)$-gravity with 
two $U(1)$-gauge bosons 
and a dilaton.
In particular, we will consider the action 
\begin{align} 
S=\int d^{4}x \sqrt{-g}[-f(R)+2\partial^{\mu}\phi \partial_{\mu}\phi 
+2\nabla_{\mu}\phi\nabla_{\nu}\phi-e^{-2\phi}(2F_{\mu\lambda}F_{\nu\delta}g^{\lambda\delta}-\frac{1}{2}g_{\mu\nu}F^{2})]\nonumber \\
\end{align}
where 
$$F_{\mu\nu}=\partial_{\nu}A_{\mu}-\partial_{\mu}A_{\nu},\,\,\,\,\tilde{B}_{\mu\nu}=\partial_{\nu}\tilde{B}_{\mu}-\partial_{\mu}\tilde{B}_{\nu}$$
and 
$A_{\mu},B_{\mu}$ are gauge bosons of $U(1)\times U(1)$, 
we conveniently use unit 
$2\kappa_{(4)}=1$, where $\kappa_{(4)}$ is the four-dimensional gravitational coupling
(coming from the Kaluza-Klein reduction of the ten-dimensional gravitational coupling). 
The action Eq.(1) comes 
 from the $SO(4)$, $d=4$, $\mathcal{N}=4$ supergravity
 and it is formulated in the Einstein-frame,
 with an opportune and understood redefinition 
 of the dilaton field.

The Equations of Motion are 
\be \label{eomA}
0=\nabla_{\mu}(e^{-2\phi}F^{\mu\nu}) \, ,
\ee
\be \label{eomB}
0=\nabla_{\mu}(e^{2\phi}\tilde{G}^{\mu\nu})  \, , 
\ee
\be \label{eomC}
0=\nabla^{2}\phi-\frac{1}{2}e^{-2\phi}F^{2}+\frac{1}{2}e^{2\phi}\tilde{G}^{2} \, , 
\ee
\be \label{eomD}
0=f_{R}(R)R_{\mu\nu}+\frac{1}{2}(Rf_{R}-f(R))g_{\mu\nu}
-\nabla_{\mu}\nabla_{\nu}f_{R}(R)+g_{\mu\nu}\partial^{2} f_{R}(R)
\ee
$$+2\nabla_{\mu}\phi\nabla_{\nu}\phi-e^{-2\phi}(2F_{\mu\lambda}F_{\nu\delta}g^{\lambda\delta}-\frac{1}{2}g_{\mu\nu}F^{2}) 
-e^{2\phi}(2\tilde{G}_{\mu\lambda}\tilde{G}_{\nu\delta}g^{\lambda\delta}-\frac{1}{2}g_{\mu\nu}\tilde{G}^{2})\, .$$

A solution of these equations is 
\begin{equation} \label{metric}
ds^{2}=e^{2U}dt^{2}-e^{-2U}dr^{2}-R^{2}d\Omega 
\end{equation}
$$e^{2\phi}=e^{2\phi_{0}}\frac{r+\Sigma}{r-\Sigma},\,\,\,F=\frac{Qe^{\phi_{0}}}{(r-\Sigma)^{2}}dt \wedge dr\,,$$
$$\tilde{G}=\frac{Pe^{-\phi_{0}}}{(r+\Sigma)^{2}}dt \wedge dr,\,\,\,e^{2U}=\frac{(r-r_{+})(r-r_{-})}{R^{2}}\,,$$
$$R^{2}=r^{2}-\Sigma^{2},\,\,\,\,\Sigma=\frac{P^{2}-Q^{2}}{2M},\,\,\,\, r_{\pm}=M\pm r_{0}\,,$$
$$r_{0}^{2}=M^{2}+\Sigma^{2}-P^{2}-Q^{2}=M^{2}+\Sigma^{2}-e^{-2\phi_{0}}P_{m}^{2}-e^{-2\phi_{0}}Q^{2}_{el}\,.$$

The solutions depend on independent parameters $M,Q,P,\phi_{0}$.
$M$ is the BH mass, $\phi_{0}$ is the asymptotic value of 
the dilaton field. 
$Q_{el}=e^{\phi_{0}Q}$ is the F-field electric charge, 
while 
$P_{m}=e^{\phi_{0}}P$ is the G-field magnetic charge
(electric charge of $\tilde{G}$).

These equations imply the relation 
$$Cf_{R}(R_{0})=q^{2}\equiv \sqrt{Q^{2}+P^{2}}=e^{-\phi_{0}}\sqrt{Q_{el}^{2}+P_{m}^{2}}$$
where $C$ is an integration constant. 

In the case of an extremal dyonic black hole, the
metric can be conveniently rewritten as \cite{Nojiri:2014jqa}
$$ds^{2}=\frac{M^{2}}{{\rm cosh}^{2}x}(d\tau^{2}-dx^{2})+M^{2}d\Omega^{2}$$

This suggests the ansatz 
$$ds^{2}=M^{2}e^{2\rho(x,\tau)}(d\tau^{2}-dx^{2})+M^{2}e^{-2\varphi(x,\tau)}(d\tau^{2}-dx^{2})d\Omega^{2}$$
and the gravitational EoM can be rewritten as 
\begin{align} 
 0=-(-\ddot{\rho}+2\ddot{\varphi}+\rho''-2\dot{\phi}^{2}-2\rho'\varphi'-2\dot{\rho}\dot{\varphi})f_{R} 
+ \frac{M^{2}}{2}e^{2\rho}f+\frac{\partial^{2}}{\partial \tau^{2}}f_{R} \nonumber\\
 -\rho'\frac{\partial}{\partial x}f_{R}+\dot{\rho}\frac{\partial}{\partial \tau}f_{R} +\frac{q^{2}M^{2}e^{2\rho}}{2}    
+e^{2\varphi}\left[-\frac{\partial}{\partial \tau}\left(e^{-2\varphi}\frac{\partial f_{R}}{\partial \tau} \right) +\frac{\partial}{\partial x} \left(e^{-2\varphi}\frac{\partial f_{R}}{\partial x} \right)    \right]\, ,
\end{align}

\begin{align} 
0=\frac{-M^{2}}{2}e^{2\rho}f-\left(\ddot{\rho}+2\varphi''-\rho''-2\varphi'^{2}-2\rho'\varphi'-2\dot{\rho}\dot{\varphi} \right)f_{R}  \nonumber\\
-\frac{q^{2}M^{2}e^{2\rho}}{2}+\frac{\partial^{2}}{\partial x^{2}}f_{R}-\dot{\rho}\frac{\partial f_{R}}{\partial \tau}
-\rho' \frac{\partial f_{R}}{\partial x}-e^{2\varphi}\left[-\frac{\partial}{\partial \tau}\left( e^{-2\varphi}\frac{\partial f_{R}}{\partial \tau} \right) +\frac{\partial}{\partial x}\left(e^{-2\varphi}\frac{\partial f_{R}}{\partial x} \right) \right]  
\end{align}

\begin{equation} \label{eq3}
0=-(2\dot{\varphi}'-2\varphi'\dot{\varphi}-2\rho'\dot{\varphi}-2\dot{\rho}\varphi')f_{R}+\frac{\partial^{2}f_{R}}{\partial \tau \partial x}-\dot{\rho}\frac{\partial f_{R}}{\partial x}-\rho'\frac{\partial f_{R}}{\partial \tau}
\end{equation}

\begin{align} 
0=-2M^{2}e^{-2\varphi}f-e^{-2(\rho+\varphi)}(-\ddot{\varphi}+\varphi''+2\varphi'^{2}+2\dot{\varphi}^{2})f_{R}
+f_{R}  
+e^{-2(\rho+\varphi)}\left( \dot{\varphi}\frac{\partial f_{R}}{\partial t}-\varphi' \frac{\partial f_{R}}{\partial x}   \right)+\frac{q^{2}M^{2}e^{2\rho}}{2}
\nonumber  \\
-e^{-2\rho}\left[  -\frac{\partial}{\partial \tau}\left( e^{-2\varphi}\frac{\partial f_{R}}{\partial \tau}\right)+\frac{\partial}{\partial x} \left( e^{-2\varphi}\frac{\partial f_{R}}{\partial x}\right)  \right]
\end{align}

Now, let us consider perturbations around the background extremal solution as 
\begin{equation}
\label{rhovarphi}
\rho=-{\rm ln}({\rm cosh}\, x)+\delta \rho, \,\,\,\varphi=\delta \varphi
\end{equation}
The perturbed EoM are 
\begin{align} 
0=\frac{f_{R}(R_{0})+2M^{-2}f_{RR}(R_{0})}{2}\delta R  
-f_{R}(R_{0})M^{-2}{\rm cosh}^{2}x (-\delta \ddot{\rho}+2\delta \ddot{\varphi}+\delta \rho''+2{\rm tanh}\, x\, \delta \phi')
\nonumber  \\
-2f_{R}(R_{0})M^{-2}\delta \rho+f_{RR}(R_{0})M^{-2}{\rm cosh}^{2}x({\rm tanh}\,x\, \delta R'+\delta R'')
\end{align}

\begin{align} 
0=-\frac{f_{R}(R_{0})+2M^{-2}f_{RR}(R_{0})}{2}\delta R +2f_{R}(R_{0})M^{-2}\delta \rho  
-f_{R}(R_{0})M^{-2}{\rm cosh}^{2}\,x\,(\delta \ddot{\rho}+2\delta \varphi''-\delta \rho''+2{\rm tanh}\, x\,\delta \varphi')
 \nonumber \\
 +f_{RR}(R_{0})M^{-2}{\rm cosh}^{2}\,x\,({\rm tanh}\, x\,\delta R'+\delta \ddot{R})
\end{align}

\begin{align} 
0=-2(\delta \dot{\varphi}'+{\rm tanh}\, x\, \delta \dot{\varphi})
+\frac{f_{RR}(R_{0})}{f_{R}(R_{0})}(\delta \dot{R}'+{\rm tanh}\, x\, \delta \dot{R})
\end{align}

\begin{align} 
0=-\frac{f_{R}(R_{0})+2M^{-2}f_{RR}(R_{0})}{2}\delta R
-2M^{-2}f_{R}(R_{0})\delta \varphi
-f_{R}(R_{0})M^{-2}\,{\rm cosh}^{2}\,x\, (-\delta \ddot{\varphi}+\delta \varphi'') \nonumber\\
-f_{RR}(R_{0})M^{-2}{\rm cosh}^{2}\,x\,(-\delta \ddot{R}+\delta R'')
\end{align}

A convenient parametrization of perturbations is 
\begin{equation}
\label{variations}
\delta \rho=\rho_{0}{\rm cosh}\, \omega \tau {\rm cosh}^{\beta}x,\,\,\,\,\delta \varphi=\varphi_{0}{\rm cosh}\, \omega \tau {\rm cosh}^{\beta}x
\end{equation}
where $\rho_{0},\phi_{0},\beta$ are arbitrary constants. 

Solving EoM, 
we find conditions 
\begin{equation}
\label{omegab}
\omega^{2}=\beta^{2}
\end{equation}
and 
\begin{equation}
\label{beta}
\beta=\beta_{\pm}=\frac{1}{2}\left[1\pm \sqrt{1-\frac{4}{3}M^{2}\left(\frac{f_{R}(R_{0})}{f_{RR}(R_{0})} \right)} \right]
\end{equation}
from 
\begin{equation}
\label{box}
\partial^{2} \delta \varphi=[\beta^{2}+\beta(\beta-1){\rm cosh}^{-2} \,x-\omega^{2}]\delta \varphi
\end{equation}
Let us note that 
$\beta$ has always a Real part which is positive, implying 
exponential instabilities.
In particular, for $\phi_{0}<0$ the antievaporation phase is obtained
while
$\phi_{0}>0$ corresponds to the evaporation. 
Hence, this is not enough to demonstrate that the extremal solution is unstable. 
So that, we show the numerical solution of the horizon radius 
obtained by 
EoM perturbed up to the second order in $\delta \rho,\delta \phi$.
Finally, we claim that a similar analysis in the case of the $SU(4)$-inspired 
model (despite of $SO(4)$ gauge group) leads to the same kind of instabilities,
as can be easily checked \footnote{
We mention that some solutions in other extended theories of gravity 
have also geodetic instabilities \cite{Addazi:2014mga}. 
}.

\section{Evaporation, Anti-evaporation and Hawking's radiation}

In this section, we will discuss the suppression of
Bekenstein-Hawking radiation in $f(R)$-gravity and $f(T)$-gravity.

\subsection{Path integral approach in $f(R)$-gravity}

In general,
the path integral over all Euclidean metrics and matter fields $\phi_{i},\psi_{j},A^{\mu}_{k},..$
is 
\be \label{formaly}
Z_{E}=\int \mathcal{D}g\mathcal{D}\phi_{i}\mathcal{D}\psi_{j}\mathcal{D}A^{\mu}_{k}e^{-I[g,\phi_{i},\psi_{j},A^{\mu}_{k},...]}
\ee
where $g$ the euclidean metric tensor. 
In Semiclassical General Relativity, the
leading terms in the action are
\be \label{I}
I_{E}=-\int_{\Sigma}\sqrt{g}d^{4}x\left(\mathcal{L}_{m}+\frac{1}{16\pi}R\right)+\frac{1}{8\pi}\int_{\partial \Sigma}\sqrt{h}d^{3}x(K-K^{0})
\ee
where $\mathcal{L}_{m}$ is the matter lagrangian 
$$\mathcal{L}_{m}=\frac{Y^{ii'}}{2}g_{\mu\nu}\partial \phi^{i\mu}\partial \phi^{i'\nu}+...$$
 $K$ the trace of the curvature induced on the boundary $\partial \Sigma$
of the region $\Sigma$ considered, $h$ is the metric induced on the boundary $\partial \Sigma$,
$K^{0}$ the trace of the curvature induced imbedded in flat space.
The last term is a contribution from the boundary.
We consider infinitesimal perturbations of matter and metric
as $\phi=\phi_{0}+\delta \phi$,
$A=A^{0}+ \delta A$, (...)
and $g=g_{0}+\delta g$, 
so that 
$$I[\phi,A,...,g]=I[\phi_{0},A_{0},..g_{0}]+I_{2}[\delta \phi,\delta A,...\delta g]+higher\,orders$$
$$I_{2}[\delta \phi,\delta A,..,\delta g]=I_{2}[\delta \phi,\delta A,...]+I_{2}[\delta g]$$
\be \label{Z}
log Z=-I[\phi_{0},A^{0},...,g_{0}]+log \int \mathcal{D}\delta \phi \mathcal{D}\delta A(...) \mathcal{D}\delta g e^{-I_{2}[\delta g,\delta \phi,\delta A,...]}
\ee

In an Euclidean Schwarzschild solution, 
the metric has a time dimension compactified on a circle $S^{1}$, with periodicity $i\beta$,
and 
$$\beta=T^{-1}=8\pi M$$
 $T,M$ BH temperature and mass. 
The Euclidean S. metric has the form 
\be \label{form1}
ds_{E}^{2}=\left(1-\frac{2M}{r}\right)d^{2}\tau+\left(1-\frac{2M}{r}\right)dr^{2}+r^{2}d\Omega^{2}
\ee
A convenient change of coordinates 
$$x=4M\sqrt{1-\frac{2M}{r}}$$
leads to 
\be \label{ES}
ds_{E}^{2}=\left(\frac{x}{4M}\right)^{2}+\left(\frac{r^{2}}{4M^{2}}\right)^{2}dx^{2}+r^{2}d\Omega^{2}
\ee
Eq.\ref{ES} has not more a (mathematical) singularity in $r=2M$. 
The boundary 
$\partial \Sigma$ is $S^{1}\times S^{2}$
with $S_{2}$ with conveniently fixed radius $r_{0}$
The path integral becomes  
a partition function of a (canonical) ensamble, 
with an euclidean time related to the temperature $T=\beta^{-1}$.
The leading contribution to the path integral is 
\be \label{dom}
Z_{ES}= e^{-\frac{\beta^{2}}{16\pi}}
\ee
Contributions to this term are only coming 
from surface terms in the gravitational action,
{\it i.e} bulk geometry does not contribute to Eq.\ref{dom}. 

The average energy (or internal energy) is 
\be \label{average}
\langle E \rangle=-\frac{d}{d\beta}(log Z)=\frac{\beta}{8\pi} 
\ee
On the other hand, the free energy $F$ is related to $Z$
as 
\be \label{usuallyd}
F=-T\rm log Z
\ee
Finally the entropy is 
\be \label{SFU}
S=\beta(F-\langle E \rangle)
\ee
As a consequence, Bekeinstein-Hawking radiation can be related to the partition function as follows: 
\be \label{S1}
S=\beta(log Z-\frac{d}{d\beta}(log Z))=\frac{\beta^{2}}{16\pi}=\frac{1}{4}A
\ee

In $f(R)$-gravity, we can reformulate an Euclidean approach. 
Through a conformal transformation, we can be more conveniently remapped
$f(R)$-gravity to a scalar-tensor theory.
The new relevant action in semiclassical regime has a form
\be \label{NRA}
I=-\frac{1}{16\pi}\int_{\Sigma} d^{4}x\sqrt{g}\left(
f(\phi)+f'(\phi)(R-\phi)\right)-\frac{1}{8\pi}\int_{\partial \Sigma}d^{3}x\sqrt{h}f'(\phi)(K-K_{0})
\ee
that can be remap to the corresponding $f(R)$-gravity action as
\be \label{PluggingBack}
I=-\frac{1}{16\pi}\int d^{4}x\sqrt{-g}f(R)-\frac{1}{8\pi}\int d^{4}x\sqrt{h}f'(R)(K-K_{0})\, . 
\ee
Let us assume a generic spherical symmetric static solution for $f(R)$-gravity 
with an Euclidean periodic time $\tau \rightarrow \tau+\beta$ 
where $\beta=8\pi M$,
\be \label{Euc}
ds_{E}^{2}=J(r)d\tau^{2}+J(r)^{-1}dr^{2}+r^{2}d\Omega^{2}\, .
\ee
As in GR, the leading contribution is zero from the bulk geometry.
But the boundary term has a non-zero contribution.
One can evaluate the boundary integral considering 
suitable surface $\partial \Sigma$. 
In this case the obvious choice is a $S_{2}\times S_{1}$ surface with 
with radius $r$ of $S_{2}$. 
We obtain 
\be \label{KKz}
\int_{\partial \Sigma}d^{3}x\sqrt{h}f'(R)(K-K_{0})=f'(R_{0})\int_{\partial \Sigma}d^{3}x\sqrt{h}(K-K_{0})=8\pi\beta r-12\pi\beta M-8\pi\beta r\sqrt{1-\frac{r_{S}}{r}}\, , 
\ee
where $r_{S}=2M$ and $R_{0}$ is the scalar curvature of the classical black hole background,
In the limit of $r\rightarrow \infty$, the resulting action, partition function and entropy are
\be \label{ActionFinal}
I=f'(R_{0})\beta^{2},\,\,\,\,\,Z_{E}=e^{-f'(R_{0})\beta^{2}},\,\,\,\,\,S=16\pi f'(R_{0})\frac{A}{4}\, . 
\ee 
The same result was also found in \cite{Dyer:2008hb}.
This result seems in antithesis with our statements in the introduction:
Eqs.(\ref{ActionFinal}) leads to a B.H.-like radiation. 
In fact, as mentioned, a Nariai solution is nothing but 
a Schwarzschild-de Sitter one
with $J(r)=1-J(r)_{Schwarzschild}-\frac{\Lambda}{3}r^{2}$
, with a black hole 
radius $r\simeq H^{-1}$ (limit of BH mass $M\rightarrow \frac{1}{3}\Lambda^{-1/2}$),
with mass scale $\mathcal{M}=\Lambda$. 
However, result (\ref{ActionFinal}) is based on a strong assumption on the metric (\ref{Euc}):
it is assumed that the gravitational action will not lead to a dynamical evolution. 
For example, in Narai solution obtained by Nojiri and Odintsov in 
$f(R)$-gravity, $J(r,t)$ is also a function of time: the mass parameter 
is a function of time $r_{S}(t)$. As a consequence, the analysis performed
here is not valid.

As a consequence, the result got in this section has to be considered 
with caution: Eq.\ref{ActionFinal} can be applied if and only if one has a spherically symmetric stationary and static solution of $f(R)$-gravity. 

Let us also comment that the same entropy in (\ref{ActionFinal})
can be obtained by the Wald entropy charge integral. 
The Wald entropy is 
\be \label{WE}
S_{W}=-2\pi \int_{S^{2}} d^{2}x\sqrt{-h^{(2)}}\left( \frac{\delta \mathcal{L}}{\delta R_{\mu\nu\rho\sigma}}\right)_{S^{2}}\hat{\epsilon}_{\mu\nu}\hat{\epsilon}_{\rho\sigma}=\frac{A}{4G_{eff}}\, , 
\ee
where $\hat{\epsilon}$ is the antisymmetric binormal vector 
to the surface $S^{2}$
and 
\be \label{Geff}
(2\pi G_{eff})^{-1}=-\left(\frac{\delta \mathcal{L}}{\delta R_{\mu\nu\rho\sigma}}\right)_{S^{2}}\hat{\epsilon}_{\mu\nu}\hat{\epsilon}_{\rho\sigma}\, , 
\ee
leading to $G_{eff}=G/f'(R_{0})$ \cite{Briscese:2007cd}.

However, again, this result can be applied if and only if the spherical symmetric solution is static.
As argument in section 2, this is not the case of Nariai BHs in $f(R)$-gravity. 

Let us argument on the non-applicability of these results in dynamical cases. 
The euclidean path integral approach is supposing an Euclidean black hole 
inside an ideal box, in thermal equilibrium with it. However, thermodynamical 
limit can be applied only for systems in equilibrium, so that a statistical 
mechanics approach can be reasonable considered. 
But a dynamical space-time inside a box
is in general an out-of-equilibrium system. 
Infact, in the next section, we will show a simple argument leading to the conclusion that
Bekenstein-Hawking evaporation is suppressed by the increasing of the Nariai's horizon 
in $f(R)$-gravity. A thermal equilibrium at $T_{B.H.}$ in an external 
ideal box will never be approached by a dynamical Nariai black hole. 

\subsection{Bekeinstein-Hawking radiation is turned off}

Let us consider a Bekestein-Hawking pair in a dynamical horizon. These are created nearby BH horizon
and they become real in the external gravitational background. 
Now, one of this pair can pass the horizon as a quantum tunnel effect,
with a certain rate $\Gamma_{bh}$. However, the horizon is displacing outward 
the previous radius because of antievaporation effect. 
As a consequence, the Bekestein-Hawking pair will be trapped in the Black hole interior, 
 in a space-like surface $\mathcal{A}_{space-like}$. From, such a space-like surface, 
a tunnel effect of one particle is impossible. 
As a consequence, the only way to escape is 
if $\Gamma_{bh}^{-1}<\Delta t$, where 
$\Delta t$ is the minimal effective time scale
(from an external observer in a rest frame)
from a $\mathcal{A}_{time-like}\rightarrow \mathcal{A}_{space-like}$
transition -
from a surface on the bh horizon $\mathcal{A}_{time-like}$ to a surface inside the bh horizon $\mathcal{A}_{space-like}$.
However, $\Delta t$ can also be infinitesimal, of the order of $\lambda$,
where $\lambda$ is the effective separation scale between 
the B.H. pair. In fact, defining $\Delta r$ as the radius increasing 
with $\Delta t$, it is sufficient $\Delta r>\lambda$ in order to 
"eat" the Bekestein-Hawking pair in the space-like interior. 
But, for black holes with a radius $r_{S}>>l_{Pl}$,
the tunneling time is expected to be $\Gamma_{bh}^{-1}>>>\Delta t$. 
As a consequence, a realistic Bekenstein-Hawking emission is impossible for 
non-planckian black holes. 
The same argument can be iteratively applied during all the 
evolution time and the external horizon. 
That Bekestein-Hawking radiation cannot be emitted by a space-like surface was 
rigorously proven in \cite{Ellis:2013oka,Firouzjaee:2014zfa,Firouzjaee:2015bqa},
with tunneling approach, eikonal approach,
and Hawking's original derivation with 
Bogoliubov coefficients.

Let us consider this situation from the energy conservation 
point of view. 
In stationary black holes, as in Schwarzschild in GR,
the BH horizon is necessary a Killing bifurcation surface.
In fact, one can define two Killing vector fields 
for the interior and the exterior of the BH. 
In the exterior region, the Killing vector 
$\zeta^{\mu}$ is time-like, 
while in the interior is space-like. 
This aspect is crucially connected with particles energies: 
the energy of a particle is $E=-p_{\mu}\zeta^{\mu}$,
where $p^{\mu}$ is the 4-momentum of the particle. 
As a consequence, energy is always $E>0$ outside the horizon. 
while $E<0$ inside the horizon. In the Killing horizon 
a real particle creation is energetically possible.
On the other hand, in the dynamical case, 
 it is not possible to define a conserved energy of a particle $E$ for a dynamical space-time, {\it i.e.} it is not possible to define a Killing vector field for time translation in a dynamical space-time.
As discussed above, the Bekeinstein-Hawking particle-antiparticle pair 
will be displaced inside the horizon in a space-like region. 
The creation of a real particle from a space-like
region is a violation of causality. 
In fact, it is an acausal exchange of energy,
{\it i.e} of classical information. 
In fact, a particle inside the horizon is inside a
light-cone with a space-like axis.   

As shown in \cite{Ellis:2013oka}, 
one can distinguish marginally
outer trapped 3-surface \footnote{We will remind at the end of this section the definition of null trapped surface, as well as those ones of marginally outer and marginally inner trapped surfaces.}
emitting Hawking's pair (timelike surface),
from the outer non-emitting one (space-like).
Let us consider the null or optics Raychaduri equation 
for null geodesic congruences: 
\be \label{theta}
\dot{\hat{\theta}}=-\hat{\theta}^{2}-2\hat{\sigma}_{ab}\sigma^{ab}+\hat{\omega}_{cd}\hat{\omega}^{cd}-R_{\mu\nu}k^{\mu}k^{\nu}
\ee
where the hats indicate that the expansion, shear, twist and vorticity are defined for the transverse directions.
The Ricci tensor encodes the dynamical proprieties of $f(R)$-gravity EoM. 
Let us also specify that $\dot{\hat{\theta}}=\frac{\partial}{\partial \lambda}\hat{\theta}$, where $\lambda$ is the affine parameter,
while $k^{a}$ is $k^{a}=\frac{dx^{a}}{d\lambda}$, with $k^{2}=0$, and $\hat{\theta}=k^{a}_{;a}$
also defined as the relative variation of the cross sectional are
$$\hat{\theta}=2\frac{1}{A}\frac{dA}{d\lambda}$$

From (\ref{theta}) one can define an emitting marginally outer 2-surface $\mathcal{A}_{time-like}$
and the non-emitting inner 2-surface $\mathcal{A}_{space-like}$. Let us call the divergence of the outgoing null geodesics 
$\hat{\theta}_{+}$ in a $S^{2}$-surface.
 With the increasing of the  black hole gravitational field, 
$\hat{\theta}_{+}$ is decreasing (light is more bended). 
On the other hand, the divergence
of ingoing null geodesics is $\hat{\theta}_{-}<0$ 
everywhere, while $\hat{\theta}_{+}>0$ for $r>2m$ in Schwarzschild. 
The marginally outer trapped 2-surface $\mathcal{A}^{2d}_{space-like}$
is rigorously defined as a space-like 2-sphere with
\be \label{theta}
\hat{\theta}_{+}(\mathcal{A}_{space-like}^{2d})=0\, . 
\ee
As mentioned above, in a Schwarzschild BH
the radius of the $S^{2}$-sphere $\mathcal{A}_{space-like}^{2d}$
is exactly equal to the Schwarzschild radius. 
As a consequence, $S^{2}$-spheres with smaller radii 
than $r_{S}=2M$ will be trapped surfaces (TS) with 
$\theta(\mathcal{A}^{2d}_{TS})<0$.

From the 2d definition, one can construct a generalized 
definition for 3d surfaces. The dynamical horizon 
is a marginally outer trapped 3-Surface. 
It is foliated by marginally trapped 2d surfaces. 
In particular, a dynamical horizon if it can be 
foliated by a chosen family of $S^{2}$ 
with $\theta_{(n)}$ of one null normal $m_{a}$
vanishing while $\theta_{n\neq m}<0$
 for each $S^{2}$. 
In particular, one can distinguish among 
an emitting marginally outer trapped 3-surface 
$\mathcal{A}_{time-like}^{3d}$
and a non-emitting one 
$\mathcal{A}_{time-like}^{3d}$
by their derivative of $\hat{\theta}_{m}$
 with respect to 
 an ingoing null tangent vector $n_{a}$. 


\be \label{fds}
\hat{\theta}_{m}(\mathcal{A}^{3d}_{time-like})=0,\,\,\,\,\,\,\partial \hat{\theta}_{m}(\mathcal{A}^{3d}_{time-like})/\partial n^{a}>0\, ,
\ee
while the non-emitting one is define as 
\be \label{sds}
\hat\theta_{m}(\mathcal{A}^{3d}_{space-like})=0,\,\,\,\,\,\,\partial \hat{\theta}_{m}(\mathcal{A}^{3d}_{space-like})/\partial n^{a}<0\, . 
\ee

Now, armed with these definitions,  let us demonstrate that the antievaporation will displace 
the emitting marginally trapped 3-surface to 
a non-emitting space-like 3-surface. 
We can consider the Raychaudhuri equation associated to our problem.
Let us suppose an initial condition $\theta(\bar{\lambda})>0$
with $\bar{\lambda}$ an initial value of the affine parameter
$\lambda$. In the antievaporation phenomena, 
the null Raychauduri equation is bounded 
as 
\be \label{boundR}
\frac{d\hat{\theta}}{d\lambda}<-R_{ab}k^{a}k^{b} \, . 
\ee
Let us consider such an equation for an infinitesimal 
$\Delta t$, so that we can expand the Schwarzschild radius 
$$r_{S}=\frac{1}{\mathcal{M}}e^{-\phi_{0}}-\frac{1}{\mathcal{M}}\beta^{2}e^{-\phi_{0}}\phi_{0}t^{2}+\frac{1}{6\mathcal{M}}\beta^{4}e^{- \phi_{0}}\phi_{0}(-2+3\phi_{0})t^{4}+O(t^{5})$$
and we can consider only the first 0th leading term. 
For any $\lambda>\bar{\lambda}$, 
$R_{ab}k^{a}k^{b}>C>0$, where $C$ is a constant associated to the 0th leading order of $R_{ab}k^{a}k^{b}$ with time. 
As a consequence, $\hat{\theta}$ is bounded as
\be \label{rett}
\hat{\theta}(\lambda)<\hat{\theta}(\lambda)+C(\lambda-\bar{\lambda})
\ee
leading to 
$\hat{\theta}(\lambda)<0$ for $\lambda>\lambda_{1}+\hat{\theta}_{1}/C$, 
where $\lambda_{1},\hat{\theta}_{1}$ are defined in a characteristic time $t_{1}$. 
As a consequence, even for a small $\Delta t$,
a constant 0th contribution coming from antievaporation
will cause an extra effective focusing term in the Raychauduri equation. 
On the other hand, the dependence of the extra focusing term 
on time is exponentially growing. 
This formalizes the argument given above. 
As a consequence, an emitting marginally trapped 3-surface will 
exponentially evolve to a non-emitting marginally one.
Bekenstein-Hawking emission are completely suppressed by this dynamical evolution
because of space-like surface cannot emit thermal Bekenstein-Hawking radiation, mixed states
\footnote{Solutions of Raychauduri equations are strictly related to energy conditions.
In $f(R)$-gravity, energy conditions like null energy condition, are generically not satisfied \cite{Albareti:2012va,Mimoso:2014ofa}.}.

Now let us consider the Raychaudhuri equation in $f(T)$-gravity \cite{Cai:2015emx}:
\be \label{theta}
\dot{\hat{\theta}}=-\frac{1}{3}\hat{\theta}^{2}-2\hat{\sigma}_{\mu\nu}\sigma^{\mu\nu}+\hat{\omega}_{\mu\nu}\hat{\omega}^{\mu\nu}-R_{\mu\nu}U^{\mu}U^{\nu}
-\tilde{\nabla}\tilde{a}-2U^{\nu}T_{\mu\nu}^{\sigma}\left(\frac{1}{3}h_{\sigma}^{\mu}\tilde{\theta}+\tilde{\sigma}_{\sigma}^{\mu}+\tilde{\omega}_{\sigma}^{\mu}-U_{\sigma}\tilde{a}^{\mu} \right)\,,
\ee
$\hat{\theta},\hat{\sigma},\hat{\omega}$ are
the expansion, shear, twist, vorticity and acceleration in $f(T)$-gravity. 
In general,  $\hat{\theta},\hat{\sigma},\hat{\omega}$ will corrected by the torsion as: 
\be \label{q1}
\tilde{\theta}=\theta_{(GR)}-2T^{\rho}U_{\rho} \, ,
\ee
\be \label{q2}
\tilde{\sigma}_{\mu\nu}=\sigma_{(GR)\mu\nu}+2h_{\mu}^{\rho}h_{\nu}^{\sigma}K_{(\rho \sigma)}^{\lambda}U_{\lambda}\, ,
\ee
\be \label{q3}
\tilde{\omega}_{\mu\nu}=\omega_{(GR)\mu\nu}+2h_{\mu}^{\rho}h_{\nu}^{\sigma}K_{[\rho \sigma]}^{\lambda}U\, , 
\ee
\be \label{q4}
\tilde{a}_{\rho}=a_{\rho(GR)}+U^{\mu}K_{\mu\rho}^{\sigma}U_{\sigma}\, ,
\ee
where is the four velocity and 
$$\tilde{\nabla}_{\mu} U_{\nu}=\tilde{\sigma}_{\mu\nu}+\frac{1}{3}h_{\mu\nu}\tilde{\theta}+\tilde{\omega}_{\mu\nu}-U_{\mu}\tilde{a}_{\nu}$$
$\dot{\hat{\theta}}=\frac{\partial}{\partial \lambda}\hat{\theta}$, where $\lambda$ is the affine parameter. 
In the optical null case, 
and $U^{a}=k^{a}$ is $k^{a}=\frac{dx^{a}}{d\lambda}$, with $k^{2}=0$, and 
$$\hat{\theta}=k^{a}_{;a}=2\frac{1}{\Sigma}\frac{d\Sigma}{d\lambda}\,.$$
We can define an emitting marginally outer 2-surface $\Sigma_{time-like}$
and the non-emitting inner 2-surface $\Sigma_{space-like}$. 

The marginally outer trapped 2-surface $\Sigma^{2d}_{space-like}$
has a topology of space-like 2-sphere with the condition
\be \label{theta}
\hat{\theta}_{+}(\Sigma_{space-like}^{2d})=0
\ee
where $\hat{\theta}_{+}$ in a $S^{2}$-surface is 
the divergence of the outgoing null geodesics. 

Let us remember that 
$\hat{\theta}_{+}$ decrease with the increasing of the gravitational field. 
$\hat{\theta}_{+}>0$ for $r>2M$ in the Schwarzschild case. 
The opposite variable is 
the divergence
of ingoing null geodesics $\hat{\theta}_{-}$, 
$\hat{\theta}_{-}<0$ 
everywhere.

The radius of the $S^{2}$-sphere $\Sigma_{space-like}^{2d}$
coincides with the Schwarzschild radius. 
$S^{2}$-spheres with smaller radii 
than $r_{S}=2M$ will be trapped surfaces (TS)
\footnote{ A trapped null surface is a set of points individuating a closed surface on which future oriented light rays are converging. In this respect, the light rays are actually moving inwards. For any compact, orientable and space-like surface, a null trapped surface can be recovered by first finding its outward pointing normal vectors, and then by studying whether the light rays directed along these latter are converging or diverging. We will say that, given a null congruence orthogonal to a space-like two-surface that has a negative expansion rate, there exists a surface that is ``trapped''. For these peculiar features, trapped null surfaces are often deployed in the definition of apparent horizon surrounding black holes.} , i.e. 
$\theta(\Sigma^{2d}_{TS})<0$.

We can generalize these topological definition for 3d surfaces.

 The dynamical horizon 
is a marginally outer trapped 3d surface. 
It is foliated by marginally trapped 2d surfaces. 
In particular, a dynamical horizon can be 
foliated by a chosen family of $S^{2}$ 
with $\theta_{(n)}$ of a null normal vector $m_a$
vanishing while $\theta_{n\neq m}<0$,
 for each $S^{2}$. 
In particular, one can distinguish among 
an emitting marginally outer trapped 3d surface 
$\Sigma_{time-like}^{3d}$
and a non-emitting one 
$\Sigma_{time-like}^{3d}$
by their derivative of $\hat{\theta}_{m}$
 with respect to 
 an ingoing null tangent vector $n_{a}$. 


\be \label{fds}
\hat{\theta}_{m}(\Sigma^{3d}_{time-like})=0,\,\,\,\,\,\,\frac{\partial \hat{\theta}_{m}(\Sigma^{3d}_{time-like})}{\partial n^{a}}>0
\ee
and  the non-emitting one is define as 
\be \label{sds}
\hat\theta_{m}(\Sigma^{3d}_{space-like})=0,\,\,\,\,\,\,\frac{\partial \hat{\theta}_{m}(\Sigma^{3d}_{space-like})}{\partial n^{a}}<0
\ee
Now, adopting these definitions,  
we 
 demonstrate that the antievaporation will transmute  
the emitting marginally trapped 3d surface to 
a non-emitting space-like 3d surface. 
We can consider the Raychaudhuri-Landau equation associated to our problem.
Let us suppose an initial condition $\theta(\bar{\lambda})>0$
with $\bar{\lambda}$ an initial value of the affine parameter
$\lambda$. In the antievaporation phenomena, 
the null Raychauduri-Landau equation is bounded 
as 
\be \label{boundR}
\frac{d\hat{\theta}}{d\lambda}<-\mathcal{R}_{ab}k^{a}k^{b}
\ee
where $\mathcal{R}_{ab}k^{a}k^{b}$ is the effective contraction of the Ricci tensor with null 4-vectors, corrected by torsion contributions:
\be \label{boundR}
\mathcal{R}_{\mu\nu}k^{\mu}k^{\nu}=R_{\mu\nu}k^{\mu}k^{\nu}+\frac{2}{3}T^{\rho}k_{\rho}-2h_{\mu}^{\rho}h_{\nu}^{\sigma}K_{(\rho \sigma)}^{\lambda}k_{\lambda}-2h_{\mu}^{\rho}h_{\nu}^{\sigma}K_{[\rho \sigma]}^{\lambda}k_{\lambda}+k^{\mu}K_{\mu\rho}^{\sigma}k_{\sigma}k^{\rho}
\ee
$$+2k^{\nu}T_{\mu\nu}^{\sigma}\left(-\frac{2}{3}h_{\sigma}^{\mu}T^{\rho}k_{\rho}
+2h_{\mu}^{\rho}h_{\nu}^{\sigma}K_{(\rho \sigma)}^{\lambda}k_{\lambda}+2h_{\mu}^{\rho}h_{\nu}^{\sigma}K_{[\rho \sigma]}^{\lambda}k_{\lambda}-k_{\sigma}k^{\mu}K_{\mu\rho}^{\sigma}k^{\rho} \right)\,,
$$

Let us consider the antievaporation case:
for  $\lambda>\bar{\lambda}$, it is
$\mathcal{R}_{ab}k^{a}k^{b}>K>0$, where $K$ is the 0-th leading order of the scalar function $\mathcal{R}_{ab}k^{a}k^{b}(t)$. 
So that
\be \label{rett}
\hat{\theta}(\lambda)<\hat{\theta}(\lambda)-K(\lambda-\bar{\lambda})
\ee
leading to 
$\hat{\theta}(\lambda)<0$ for $\lambda>\lambda_{0}+\hat{\theta}_{0}/K$, 
where $\lambda_{0},\hat{\theta}_{0}$ are defined at a  characteristic time $t_{0}$. 
For a small $\delta t$, 
a constant 0th contribution sourced by the torsion 
will cause an effective focusing term in the Raychauduri equation. 
This phenomena is exponentially growing in time. 
So that, an emitting marginally trapped 3d surface will 
exponentially evolve to a non-emitting marginally one.

Now let us consider a Bekenstein-Hawking pair in an antievaporating solution. 
They are imagined to be created in the black hole horizon 
as virtual pair. Then the external gravitational field can 
promote them to be real particles.
Then, a particle of this pair 
can quantum tunnel outside the black hole horizon
with a certain characteristic time scale $\tau_{bh}$. 
With an understood correction to the Black hole entropy formula, this conclusion seems compatible with Nariai solutions in diagonal tetrad choice. 
 Bekenstein-Hawking's calculations are performed in the limit of a static horizon 
and a black hole in thermal equilibrium with the environment.
This approximation cannot work for antievaporating black holes. 
In fact, the horizon is displacing outward 
the previous radius. 
The Bekenstein-Hawking pair will be trapped in the black hole interior, 
foliated in space-like surfaces $\Sigma_{space-like}$.
But 
from a space-like surface,
the tunneling effect of a particle is impossible: 
otherwise causality will be violated. 
As a consequence, Bekenstein-Hawking radiation requests
$\tau_{bh}<\delta t$, where 
$\delta t$ is the minimal effective time scale in the external rest frame
for a
$\Sigma_{time-like}\rightarrow \Sigma_{space-like}$ transition. 
The Bekenstein-Hawking radiation is exponentially turned off with time. 
Infact  Bekenstein-Hawking radiation cannot be emitted from a space-like surface
in all possible approaches, as proven in \cite{Ellis:2013oka,Firouzjaee:2014zfa,Firouzjaee:2015bqa}.
 
\subsection{A new radiation in non-diagonal evaporating solutions}

Now, let us comment what happens in the opposite case: evaporating solutions. 
In this case $f(T)$-gravity will source an extra anti-focalizing term in the null Raychauduri equation.
This will cause exactly the opposite transition: a null-like horizon is pushed out the black hole radius and it will become 
time-like. Defining $\delta t$ as the transition time $\Sigma_{space-like}\rightarrow \Sigma_{time-like}$, 
Bekenstein-Hawking effect will happen if $\tau_{bh}<<\delta t$. 
However, with $\delta t<\tau_{bh}$, the Bekenstein-Hawking pair is pushed-off from the black hole horizon. 
In other words, they {\it both} will be emitted from the black hole. 
They can annihilate outside the black hole producing radiation. 
Contrary to Bekenstein-Hawking radiation, unitarity is not violated in black hole formation during the gravitational collapse. 
In fact, the firewall paradox is exactly coming by from the entanglement of 
the two pairs combined by the fact the one is falling inside the interior while its twin tunnels out. 
In our case, both are emitted outwards because of evaporation effects.  
In Bekenstein-Hawking case, outgoing information is exactly copied with the interior 
information. In our case, there is not any entanglement among black hole interior 
and external environment. 
 This radiation does not introduce any new information paradoxes. 

\section{Brane-worlds instabilities }

In this section, we will study the presence of evaporation and antievaporation 
instabilities in Brane-world scenarios \cite{Addazi:2017puj}. 
Let us consider the $F(R)$-gravity theory in five dimensions;
\begin{equation}
\label{A}
S=\frac{1}{2\kappa_{5}^{2}}\int \sqrt{-g}\left[ F^{(5)}(R)+S_{m} \right] \, , 
\end{equation} 
where $\kappa_{5}$ is the five-dimensional gravitational constant and $S_{m}$ is the action of the 
matter.
The equations of motion in the vacuum are given by 
\begin{equation} 
\label{AA}
F^{(5)}_{R}(R)
\left(R_{\mu\nu}-\frac{1}{2}Rg_{\mu\nu} \right) 
=\frac{1}{2}g_{\mu\nu} \left[ F^{(5)}(R)-RF_{R}^{(5)}(R) \right]
+\left[ \nabla_{\mu}\nabla_{\nu}-g_{\mu\nu}\partial^{2} \right]F_{R}^{(5)}(R) \, ,
\end{equation}
where $F^{(5)}_{R}= d F^{(5)}/ d R$.
Especially if we assume that the metric is covariantly constant, that is, 
$R_{\mu\nu} = K g_{\mu\nu}$ with a constant $K$, we find 
\be 
\label{AA2} 
0= RF^{(5)}_R (R) - \frac{5}{2}F^{(5)} (R)\, .
\ee
We denote the solution of Eq.~(\ref{AA2}) as $R=R_0$ and define the length parameter $l$ by 
$R_{0}=20/l^{2}$.
We should note that the metric of the Schwarzschild-de Sitter solution is covariantly constant 
and given by, 
\be 
\label{B} 
ds_{SdS, (5)}^{2}= \frac{1}{h(a)}da^{2}-h(a)dt^{2}+a^{2}
d\Omega_{(3)}^{2}\, ,\quad
h(a)= 1-\frac{a^{2}}{l^{2}}-\frac{16\pi G_{(5)} M}{3 a^{2}} \, .
\ee
Here $M$ corresponds to the mass of the black hole and 
$G_{(5)}$ is defined by $8\pi G_{(5)} = \kappa_{5}^2$.
The space-time expressed by the metric (\ref{B}) has two horizons at 
\be 
\label{B1}
a^2 = a_\pm ^2 = \frac{l^2}{2}
\left\{ 1 \pm \sqrt{ 1 - \frac{64\pi G_{(5)} M}{3 l^2}} \right\} \, .
\ee
The two horizons degenerate in the limit, 
\be 
\label{B2} \frac{64\pi G_{(5)} M}{3 l^2} \to 1\, , 
\ee 
and we obtain the degenerate Schwarzschild-de Sitter (Nariai) solution.
The metric in the Nariai space-time is given by 
\begin{equation} 
\label{dsdmL} 
ds^{2} =\frac{1}{\Lambda}\left(-\sin^{2}\chi d\psi^{2}+d\chi^{2}
+d\Omega_{(3)}^2 \right) \, ,
\end{equation}
where there are the horizons at $\chi=0,\, \pi$ and $\Lambda=\frac{2}{l^2}$.
Let us perform the coordinate transformation $\chi= {\rm arccos} \zeta$, 
\be 
\label{B3} 
ds^{2}=-\frac{1}{\Lambda} \left(1-\zeta^{2} \right)d\psi^{2}
+\frac{d\zeta^{2}}{\Lambda \left(1-\zeta^{2} \right)}
+\frac{1}{\Lambda}d\Omega_{(3)}^2 \, ,
\ee
which is singular at $\zeta=\pm 1$.
By changing the coordinate $\zeta={\rm tanh} \xi$, the metric can be rewritten as, 
\be 
\label{B4} 
ds^{2}=\frac{1}{\Lambda \cosh^2 \xi} \left( - d \psi^2 + d\xi^2 \right)
+\frac{1}{\Lambda}d\Omega_{(3)}^2 \, .
\ee
We often analytically continue the coordinates by 
\be 
\label{B5} 
\psi = ix\, , \quad \zeta = i\tau\, , 
\ee 
and we obtain the following metric 
\begin{equation} 
\label{dsNa} 
ds^{2}=-\frac{1}{\Lambda \cos^{2}\tau} \left(-d\tau^{2}+dx^{2} \right) 
+\frac{1}{\Lambda}d\Omega_{(3)}^2 \, .
\end{equation}
Of course, after the analytic continuation, the obtained space is a solution of the equations 
although the topology is changed.
This expression of the metric was used in \cite{Bousso:1997wi}.

In order to consider the perturbation, we now consider the general metric in the following form, 
\begin{equation} 
\label{dse} 
ds^{2}=\e^{2\rho(x,\tau)} \left( -d\tau^{2}+dx^{2} \right)
+\e^{-2\phi(x,\tau)}d\Omega_{(3)}^2 \, ,
\end{equation}
which generalizes the Nariai metric in Eq.(\ref{dsNa})
with generic functions $\rho(x,\tau), \phi(x,\tau)$.

Then the equation of motion can be decomposed in components as 
\begin{equation} 
\label{e1} 
0=-\frac{\e^{2\rho}}{2}F^{(5)} - \left( -\ddot{\rho}
+3\ddot{\phi}+\rho''-3\dot{\phi}^{2}-3\dot{\rho}\dot{\phi}-3\rho'\phi'
\right)F^{(5)}_{R}
+\ddot{F}^{(5)}_{R} 
\end{equation}
$$ -\dot{\rho}\dot{F}^{(5)}_{R}-\rho' \left(F^{(5)}_{R} \right)'
+\e^{2\phi}\left[-\frac{\partial}{\partial \tau}\left(\e^{-2\phi}\dot{F}^{(5)}_{R}
\right)+\left(\e^{-2\phi}(F^{(5)}_{R})' \right)' \right] \, , $$

\begin{equation} 
\label{e2} 
0 =\frac{\e^{2\rho}}{2}F^{(5)}-\left(-\rho''+3\phi''+\ddot{\rho}-3\phi'^{2}
 -3\rho'\phi'-3\dot{\rho}\dot{\phi}\right)F^{(5)}_{R}+{F^{(5)}_{R}}'' 
 \end{equation}
$$-\dot{\rho}\dot{F}^{(5)}_{R}-\rho' \left(F^{(5)}_{R} \right)'
  -\e^{2\phi}\left[-\frac{\partial}{\partial
\tau}\left(\e^{-2\phi}\dot{F}^{(5)}_{R}\right)
+\left(\e^{-2\phi} \left(F^{(5)}_{R} \right)'\right)' \right] \, ,$$

\begin{equation} 
\label{e3} 
0=- \left(3\dot{\phi}'-3\phi'\dot{\phi}-3\rho'\dot{\phi}-3\dot{\rho}\phi' 
\right)F^{(5)}_{R}
+\frac{\partial^{2}F^{(5)}_{R}}{\partial x \partial \tau}
 -\dot{\rho} \left(F^{(5)}_{R} \right)'-\rho'\dot{F}^{(5)}_{R} \, , 
\end{equation}
 
\begin{equation} 
\label{e4} 
 0=\frac{\e^{-2\phi}}{2}F^{(5)}-\e^{-2(\rho+\phi)}
\left( -\ddot{\phi}+\phi''+3\dot{\phi}^{2}-3\phi'^{2}
\right)F^{(5)}_{R} -F^{(5)}_{R}
+\e^{-2(\rho+\phi)}\left(\dot{\phi}\dot{F}^{(5)}_{R}
  -\phi'{F^{(5)}_{RR}}' \right) 
  \end{equation}
$$-e^{-2\rho}\left[-\frac{\partial}{\partial
\tau}\left(\e^{-2\phi}\dot{F}^{(5)}_{R}\right)
+\left(\e^{-2\phi}{F^{(5)}_{RR}}' \right)' \right]\, ,$$
where
$F'= \frac{\partial F}{\partial x}$ and $\dot{F}=\frac{\partial F}{\partial \tau}$ 
and we have used the expressions of the curvatures (\ref{Ncurvature}) in the Appendix \ref{A1}.

We consider the perturbations at the first order around the Nariai background 
Eq.(\ref{dsNa}) with 
$R_{0}=\frac{20}{l^2}$, 
\begin{equation}\label{p1} 
0= \frac{-F^{(5)}_{R}(R_{0}) +2\Lambda F^{(5)}_{RR}(R_{0})}{2\Lambda \cos^{2}\tau}\delta R
  -\frac{F^{(5)}(R_{0})}{\Lambda \cos^{2}\tau}\delta \rho
  -F^{(5)}_{R}(R_{0}) \left(-\delta \ddot{\rho}+3\delta \ddot{\phi}
+\delta \rho'' -3\tan \tau \delta \dot{\phi}\right) 
\end{equation}
$$ -\tan \tau F^{(5)}_{RR}(R_{0})\delta \dot{R}+F^{(5)}_{RR}(R_{0})\delta R'' \, , \\ $$
\begin{equation}\label{p2} 
0= -\frac{-F^{(5)}_{R}(R_{0})
+2\Lambda F^{(5)}_{RR}(R_{0})}{2\Lambda \cos^{2}\tau}\delta R 
+\frac{F^{(5)}(R_{0})}{\Lambda \cos^{2}\tau}\delta \rho
  -F^{(5)}_{R}(R_{0})
\left(\delta \ddot{\rho}+3\delta \phi''-\delta \rho''
  -3\tan \tau \delta \dot{\phi} \right) 
  \end{equation}
$$ -\tan \tau F^{(5)}_{RR}(R_{0})\delta \dot{R}+F^{(5)}_{RR}(R_{0})\delta R''\, , \\ $$
\be\label{p3} 
0=-3F^{(5)}_{R}(R_{0}) \left(\delta \dot{\phi}'-\tan \tau \delta \phi' \right)
+F^{(5)}_{RR}(R_{0}) \left(\delta \dot{R}'-\tan \tau \delta R' \right)
\, , \\
\ee
\be\label{p4}
0= - \frac{-F^{(5)}_{R}(R_{0})
+2\Lambda F^{(5)}_{RR}(R_{0})}{2\Lambda \cos^{2}\tau }\delta R
 -\frac{F^{(5)}(R_{0})}{\Lambda \cos^{2}\tau }\delta \phi
 -F^{(5)}_{R}(R_{0})\left(-\delta \ddot{\phi}+\delta \phi'' \right) 
 \ee
$$- F^{(5)}_{RR}(R_{0})(-\delta \ddot{R}+\delta R'') \, .$$

The perturbation of the scalar curvature $\delta R$ is given in terms of $\delta \rho$ and 
$\delta \phi$ as follows, 
\begin{equation} 
\label{dR} 
\delta R=4\Lambda(-\delta \rho+\delta \phi)
+\Lambda \cos^{2}\tau(2\delta \ddot{\rho}-2\delta \rho''-6\delta \ddot{\phi}+6\delta \phi'')\,.
\end{equation}
Therefore the four equations of motions include only two 
$\delta \phi$ and $\delta \rho$, which tell that only two equations in the four equations 
 should be independent ones.

One can find that Eq.~(\ref{p3}) can be easily integrated 
\begin{equation} 
\label{C1} 
\delta R=3\frac{F^{(5)}_{R}(R_{0})}{F^{(5)}_{RR}(R_{0})}\delta \phi
+\frac{c_{1}(x)}{\cos \tau}+c_{2}(\tau)\, .
\end{equation}
Here $c_1(x)$ and $c_2(\tau)$ are arbitrary functions but because $\delta R$ should vanish 
when both of $\delta \rho$ and $\delta \phi$ vanish as seen from (\ref{dR}), we can put 
$c_1(x)=c_2(\tau)=0$.

Then, one can directly consider Eq.(\ref{dR}):
Substituting in it $\delta R(\delta \phi)$ obtained in Eq.(\ref{C1}), we find a simple equation 
\begin{equation} 
\label{E} 
\left(\partial^{2}+\frac{M^{2}}{\cos^{2}\tau}\right)\delta \phi=0 \, , \quad 
\partial^{2} \equiv - \frac{\partial^2}{\partial \tau^2} + \frac{\partial^2}{\partial x^2}\, .
\end{equation}
Here
\begin{equation}
\label{alpha}
M^{2}=\frac{1}{2}\frac{4\alpha-1}{\alpha}\, , \quad \alpha 
=\frac{4\Lambda F^{(5)}_{RR}(R_{0})}{F^{(5)}_{R} \left(R_{0} \right)} 
=\frac{F(R_{0})F_{RR}(R_{0})}{[F_{R}(R_{0})]^{2}} \, .
\end{equation}
Eq.~(\ref{E}) is nothing but a time-dependent Klein-Gordon equation for the $\delta\phi$ mode, 
with an effective oscillating mass term in time.
An explicit solution of (\ref{E}) is given by 
\be 
\label{S1} 
\delta \phi = \phi_0 \cos \left( \beta x \right) \cos^\beta \tau \, .
\ee
Here $\beta$ is given by solving the equation $M^2 = \beta \left( \beta - 1 \right)$.
The anti-evaporation corresponds to the increasing of the radius of the apparent horizon, which is 
defined by the condition, 
\begin{equation} 
\label{condition} 
\nabla \delta \phi \cdot \nabla \delta \phi=0 \, .
\end{equation}
In other words, it is imposed that the (flat) gradient of the two-sphere size is null.
By using the solution in (\ref{S1}), we find $\tan \beta x = \tan \tau$, that is, $\beta x = \tau$.
Therefore on the apparent horizon, we find 
\be\label{S2} 
\delta \phi = \phi_0 \cos^{\beta+1} \tau \, .
\ee
Because the horizon radius $r_H$ is given by $r_H = \e^{- \phi}$, we find 
\be\label{S3} 
r_H = \frac{\e^{- \phi_0 \cos^{\beta+1} \tau}}{\sqrt{\Lambda}}\, .
\ee
Then if $\beta < -1$, the horizon grows up, which corresponds to the anti-evaporation depending 
on the sign of $\phi_{0}$.
The sign could be determined by the initial condition of the perturbation.
On the other hand, it is also possible the case in which $\beta,\omega$ are complex parameters.
In this case, solutions of perturbed equations read 
\be\label{NS} 
\delta \phi={\rm Re}\left\{(C_{1}\e^{\beta t}+C_{2}\e^{-\beta t})\e^{\beta x} \right\}\, , 
\ee 
where $C_{1,2}$ are complex numbers.
$\delta \phi$ always increase in time for $C_{1}\neq 0$ because of
${\rm Re}\beta>0$.
This means that the Nariai solution is unstable also in this region of parameters.
A particular class among possible complex parameter solutions is 
\be\label{NS2} 
\delta \phi= \phi_{0}\left\{\e^{\frac{-t+x}{2}}\left(\cos \frac{\gamma(t-x)}{2}
+\frac{1}{\gamma}\sin \frac{\gamma(t-x)}{2}\right) 
+\e^{\frac{t+x}{2}}\left(\cos \frac{\gamma(t+x)}{2}
 -\frac{1}{\gamma}\sin \frac{\gamma(t+x)}{2}\right)\right\}\,,
\ee
where 
$\beta\equiv \frac{1}{2}(1+i\gamma)$ and $\gamma\equiv \pm \sqrt{\frac{2-9\alpha}{\alpha}}$.

On the horizon, the fluctuations must satisfy the condition 
$\frac{\phi_{0}^{2}}{2}\gamma^{2}\e^{x}\sin\frac{\gamma(t-x)}{2}
\sin \frac{\gamma(t+x)}{2}=0$,
which corresponds to two classes of solutions with $x=\mp t+\frac{2n\pi}{\gamma}$, 
\be
\label{NS2} 
\delta \phi=\phi_{0}(-1)^{n}\left\{ \e^{\frac{n\pi}{\gamma}}
+\e^{\mp t+\frac{n\pi}{\gamma}}\left(\cos\gamma t\mp
\frac{1}{\gamma}\sin \gamma t \right)\right\}\,, 
\ee 
which implies an oscillating horizon radius.

Let us consider a class of $F^{(5)}(R)$ models 
\begin{equation} 
\label{fRc} 
F^{(5)}(R)=\frac{R}{2\kappa^{2}}+f_{2}R^{2}+f_{0}\mathcal{M}^{5-2n}R^{n}\, .
\end{equation}
Here $f_2$ and $\mathcal{M}$ are constants with a mass dimension and $f_0$ is a dimensionless 
constant.
In this case, $\alpha$ is given by
\begin{equation}
\label{ps}
\alpha=\frac{4\Lambda
\left( 2f_{2}+n(n-1)f_{0}\mathcal{M}^{5-2n}R_{0}^{n-2} \right)}{1/2\kappa^{2}+2f_{2}R_{0} 
+nf_{0}M^{5-2n}R_{0}^{n-1}}\, .
\end{equation}
Then $\beta$ is given by
\be
\label{beta1}
\beta^2 - \beta =\frac{1}{2\alpha}(4\alpha-1) \, , \ee that is \be \label{beta2} \beta_{\pm} 
= \frac{1}{2} \left( 1 \pm \sqrt{ \frac{9\alpha-2}{\alpha}} \right) \, .
\ee
Then the condition of the anti-evaporation $\beta<-1$ (for $\phi_{0}<0$) can be satisfied only by 
$\beta_{-}$ and for $\alpha<0$.
On the other hand, for $\beta$ as a complex parameter in Eq.(\ref{NS2}), the oscillation instabilities 
are obtained for $0<\alpha<2/9$.
In this case, evaporation and antievaporation phases are iterated.

\subsection{Brane dynamics in the bulk }

We now consider the $F^{(d+1)}(R)$ gravity in the $d+1$ dimensional space-time $M$ with 
$d$ dimensional boundary $B$, whose action is given by 
\begin{equation} 
\label{FRd1} 
S=\frac{1}{2\kappa^2}\int_M d^{d+1} x \sqrt{-g} F^{(d+1)}(R)\, , 
\end{equation} 
which can be rewritten in the scalar-tensor form.
We begin by rewriting the action (\ref{FRd1}) by introducing the auxiliary field $A$ as follows, 
\begin{equation} 
\label{FRd2} 
S=\frac{1}{2\kappa^2}\int d^{d+1} x \sqrt{-g}
\left\{{F^{(d+1)}}'(A)\left(R-A\right) + F^{(d+1)}(A)\right\}\, .
\end{equation}
By the variation of the action with respect to $A$, we obtain the equation $A=R$ and by 
substituting the obtained expression $A=R$ into the action (\ref{FRd2}), we find that the action in 
(\ref{FRd1}) is reproduced.
If we rescale the metric by conformal transformation, 
\begin{equation} 
\label{JGRG22} 
g_{\mu\nu}\to \e^\sigma g_{\mu\nu}\, ,\quad \sigma = -\ln {F^{(d+1)}}'(A)\, , 
\end{equation} 
we obtain the action in the Einstein frame, 
\be\label{FRd3} 
S_E = \frac{1}{2\kappa^2}\int_M d^{d+1} x \sqrt{-g} \left(R
 - (d-1)\partial^{2} \sigma - \frac{(d-2)(d-1)}{4}\partial^\mu \sigma \partial_\mu \sigma
 - V(\sigma)\right) 
 \ee
$$= \frac{1}{2\kappa^2}\int_M d^{d+1} x \sqrt{-g} \left(R 
 - \frac{(d-2)(d-1)}{4}\partial^\mu \sigma \partial_\mu \sigma
 - V(\sigma)\right)
+ (d-1) \int_B d^d x \sqrt{-\hat g} n^\mu \partial_\mu \sigma \, , \nn$$
$$V(\sigma) =e^\sigma g\left(e^{-\sigma}\right)
 - e^{2\sigma} f\left(g\left(e^{-\sigma}\right)\right)
 = \frac{A}{{F^{(d+1)}}'(A)} - \frac{F^{(d+1)}(A)}{{F^{(d+1)}}'(A)^2}\, .$$
Here $g\left(e^{-\sigma}\right)$ is given by solving the equation 
$\sigma = - \ln {F^{(d+1)}}'(A)$ as $A=g\left(e^{-\sigma}\right)$.
By the integration of the term $\partial^{2}\sigma$, there appears the boundary term, where $n^\mu$ is 
the unit vector perpendicular to the boundary and the direction of the vector is inside.
Furthermore $\hat{g}_{\mu\nu}$ is the metric induced on the boundary, 
$\hat{g}_{\mu\nu} = g_{\mu\nu} - n_\mu n_\nu$.
The existence of the boundary makes the variational principle with respect to $\sigma$ ill-defined, 
we cancel the term by introducing the boundary action 
\begin{equation} 
\label{FRd4} 
S_B = - (d-1) \int_B d^d x \sqrt{-\hat{g}} n^\mu \partial_\mu \sigma \, .
\end{equation}
Then one may forget the boundary term,
\begin{equation}
\label{FRd5}
S_E \rightarrow S_E + S_B = \frac{1}{2\kappa^2}\int_M d^{d+1} x \sqrt{-g} 
\left(R - \frac{(d-2)(d-1)}{4}\partial^{\mu} \sigma \partial_{\mu} \sigma
 - V(\sigma)\right) \, .
\end{equation}
As is well-known, because the scalar curvature $R$ includes the second derivative term, the 
variational principle is still ill-defined in the space-time with boundary \cite{Gibbons:1976ue} 
(see also, Refs.~\cite{Chakraborty:2014xla,Chakraborty:2015bja,Nojiri:2004bx,Deruelle:2007pt}).
Because the variation of the scalar curvature with respect to the metric is given by 
\begin{equation} 
\label{EE1} 
R = -\delta g_{\mu\nu} R^{\mu\nu}
+ g^{\sigma\nu} \left(\nabla_\mu \delta{\Gamma}^\mu_{\sigma\nu}
 - \nabla_\sigma \delta \Gamma^\mu_{\mu\nu} \right) \, , 
\end{equation} 
the variation of the action with respect to the metric is given by, 
\begin{equation} 
\label{EE2} 
\delta S_\mathrm{E}= \frac{1}{2\kappa^2} \int d^{d+1} x \sqrt{-g} Q^{\mu\nu} \delta g_{\mu\nu}
+ \frac{1}{2\kappa^2} \int_B d^d x \sqrt{-\hat g}
g^{\sigma\nu} \left( - n_\mu \delta \Gamma ^\mu_{\sigma\nu}
+ n_\sigma \delta \Gamma ^\mu_{\mu\nu} \right)\, .
\end{equation}
Here the Einstein equation in the bulk is given by $Q_{\mu\nu}=0$.
Then the variational principle becomes well-defined if we add the following boundary term, 
\begin{equation} 
\label{E3} 
\tilde S_b=- \frac{1}{2\kappa^2} \int_B d^d x \sqrt{-\hat g} g^{\sigma\nu} \left(
 - n_\mu \Gamma ^\mu_{\sigma\nu}+ n_\sigma \Gamma ^\mu_{\mu\nu} \right)\, .
\end{equation}
Although the above boundary term (\ref{E3}) is not invariant under the reparametrization, because 
\begin{equation} 
\label{E4} 
\nabla_\mu n_\nu=\partial_\mu n_\nu
 - \Gamma_{\mu\nu}^\lambda n_\lambda \ ,\quad \nabla_\mu n^\nu=\partial_\mu n^\nu
+ \Gamma_{\mu\lambda}^\nu n^\lambda\, ,
\end{equation}
we find
\begin{equation}
\label{E5}
g^{\sigma\nu} \left( - n_\mu \Gamma ^\mu_{\sigma\nu}
+ n_\sigma \Gamma ^\mu_{\mu\nu} \right)
= - \partial_\mu n^\mu - 2g^{\delta\rho}\partial_\delta n_\rho
\nabla_\mu n^\mu\, ,
\end{equation}
which is just equal to $\nabla_{\mu}n^{\mu}$ on the boundary  
\cite{Chakraborty:2014xla,Chakraborty:2015bja,Nojiri:2004bx,Deruelle:2007pt,Gibbons:1976ue} . 
Therefore we can replace the boundary term (\ref{E3}) by the Gibbons-Hawking boundary term, %
\begin{equation}
\label{E6} 
S_{GH} = \frac{1}{\kappa^{2}} \int_{B} d^{d} x \sqrt{-\hat{g}}\nabla_{\mu} n^{\mu}\, .
\end{equation}

Let the boundary is defined by a function $f(x^\mu)$ as $f(x^\mu)=0$.
Then by the analogy of the relation between the electric field and the electric potential in the 
electromagnetism, we find that the vector $\left( \partial_\mu f\left( x^\mu \right) \right)$ is 
perpendicular to the boundary because 
$dx^\mu \partial_\mu f\left( x^\mu\right) = 0$ on the
boundary, which gives an expression for $n_\mu$ as 
\be 
\label{n1} 
n_\mu = \frac{ \partial_\mu f}{\sqrt{g^{\rho\sigma} \partial_\rho f \partial_\sigma f}}\, .
\ee
Then with respect to the variation of the metric, the variation of $n^\mu$ is given by 
\be 
\label{n2} 
\delta n_\mu = \frac{1}{2} \frac{ \partial_\mu f} {\left( g^{\rho\sigma} 
\partial_\rho f \partial_\sigma f\right)^{\frac{3}{2}}} \partial^\tau f \partial^\eta f 
\delta g_{\tau\eta} = \frac{1}{2} n_\mu n^\rho n^\sigma \delta g_{\rho\sigma}\, .
\ee
By using the expression in (\ref{n2}), one finds the variation of $\nabla_\mu n^\mu$ with respect to 
the metric, 
\be 
\label{n3} 
\delta \left( 2 \nabla_\mu n^\mu \right) = - 2\delta g_{\mu\nu} n^\mu n^\nu
 - n^\mu \nabla^\nu \delta g_{\mu\nu}
 - g^{\mu\nu} n_\rho \delta \Gamma^\rho_{\mu\nu}
+ n^\nu \delta \Gamma^\mu_{\mu\nu}\, .
\ee
The last two terms in (\ref{n3}) are necessary to make the variational principle well-defined but 
the second term $n^\mu \nabla^\nu \delta g_{\mu\nu}$ also may violate the variational principle.
By using the reparametrization invariance, however, we can choose the gauge condition so that 
$\nabla^\nu \delta g_{\mu\nu}=0$.

We may also add the following boundary term, 
\begin{equation} 
\label{EE3} 
S_\mathrm{BD} = \int_B d^d x \sqrt{-\hat g} \mathcal{L}_\mathrm{B} \, , 
\end{equation} 
The variation of the total action 
\begin{equation} 
\label{EE4} 
S_\mathrm{total} = S_\mathrm{E} + S_B + S_\mathrm{GH} + S_\mathrm{BD} \, , 
\end{equation} 
is given by 
\begin{equation} 
\label{EE5} 
\delta S_\mathrm{total} = \frac{1}{2\kappa^2} \int d^{d+1} x 
\sqrt{-g} Q^{\mu\nu} \delta g_{\mu\nu}+ \int_B d^d x \sqrt{-\hat g} \left[ \frac{1}{2\kappa^2}
\left( \frac{1}{2} \mathcal{K} \hat g^{\mu\nu} - \mathcal{K}^{\mu\nu} \right)
+ \frac{1}{2}T_\mathrm{B}^{\mu\nu} \right] \delta g_{\mu\nu}\, .
\end{equation}
Here we have defined the extrinsic curvature by 
$\mathcal{K}_{\mu\nu} \equiv \nabla_{\mu} n_{\nu}$ and 
$\mathcal{K} \equiv g^{\mu\nu} \mathcal{K}_{\mu\nu}$.
We also wrote the variation of $S_\mathrm{BD}$ as 
\begin{equation} 
\label{EE6} 
\delta S_\mathrm{BD} = \frac{1}{2}\int_B d^d x \sqrt{-\hat g} 
T_\mathrm{B}^{\mu\nu} \delta g_{\mu\nu}\, .
\end{equation}
Then on the boundary, we obtain the following equation, 
\begin{equation} 
\label{EE7}
0 = \frac{1}{2} \mathcal{K} \hat g^{\mu\nu} - \mathcal{K}^{\mu\nu}
+ \kappa^2 T_\mathrm{B}^{\mu\nu} \, ,
\end{equation}
which may be called the brane equation.
Especially if the boundary action $S_\mathrm{BD}$ consists of only the brane tension 
$\tilde\kappa$, 
\begin{equation} 
\label{EE8} 
S_\mathrm{B} = \frac{\tilde\kappa}{\kappa^2} \int_B d^d x \sqrt{-\hat g} \, , 
\end{equation} 
we find 
\begin{equation} 
\label{EE9}
0 = \frac{1}{2} \mathcal{K} \hat g^{\mu\nu} - \mathcal{K}^{\mu\nu}
+ \tilde\kappa g^{\mu\nu} \, ,
\end{equation}
which can be rewritten as,
\begin{equation}
\label{EE11}
0 = \frac{2}{d-2} \tilde\kappa \hat g^{\mu\nu} - \mathcal{K}^{\mu\nu} \, .
\end{equation}
If we consider the model which is given by gluing two space-time as 
in the Randall-Sundrum model \cite{Randall:1999ee,Randall:1999vf}, 
the contribution from the bulk doubles and therefore the Gibbons-Hawking term also doubles, 
\begin{equation} 
\label{EE12}
0 = \frac{2}{d-2} \tilde\kappa \hat g^{\mu\nu} - 2 \mathcal{K}^{\mu\nu} \, .
\end{equation}
Let us consider the following five-dimensional geometry, 
\begin{equation} 
\label{SAdS} 
ds^{2}_\mathrm{5}= g_{\mu\nu}dx^\mu dx^\nu = - \e^{2\rho} dt^2 + \e^{-2\rho} da^2
+ a^2 d \Omega_\mathrm{3}^2 \, .
\end{equation}
Here $d \Omega_\mathrm{3}^2 = \tilde g_{ij} dx^i dx^j$ expresses the metric of the unit sphere in 
two dimensions.
We now introduce a new time variable $\tau$ so that the following condition is satisfied, 
\begin{equation} 
\label{cd1}
 -\e^{2\rho}\left( \frac{\partial t}{\partial \tau} \right)^2
+\e^{-2\rho}\left( \frac{\partial a}{\partial \tau} \right)^2 = -1 \ .
\end{equation}
Then we obtain the following FRW metric
\begin{equation}
\label{met1}
ds^{2}_4=\tilde g_\mathrm{ij} dx^i dx^j = -d\tau ^2 +a^2 d \Omega_\mathrm{3}^2 \, .
\end{equation}
Then
\begin{equation}
\label{nmu}
n^\mu =\left( - \e^{-2\rho} \frac{\partial a}{\partial \tau},
 -\e^{2\rho} \frac{\partial t}{\partial\tau}, 0,0,0 \right)\, .
\end{equation}
Because
\begin{equation}
\label{dn}
\mathcal{K}_{ij} = \frac{\kappa}{2}\e^{4\rho} a \tilde g_{ij} \frac{dt}{d\tau}\, , 
\end{equation} 
from Eq.~(\ref{EE9}), we obtain, 
\begin{equation} 
\label{dn2} 
\e^{2\rho} \frac{dt}{d\tau} = - \frac{\tilde\kappa}{2} a \, .
\end{equation}
Using (\ref{cd1}) and defining the Hubble rate by $H= \frac{1}{a} \frac{da}{d\tau}$, one finds the 
following FRW equation for the brane, 
\begin{equation} 
\label{HH0}
H^2 = - \frac{\e^{2\rho(a)}}{a^2} + \frac{\tilde\kappa^2}{4}\, .
\end{equation}
Then in case of the Schwarzschild-de Sitter black hole, 
\begin{equation} 
\label{SdS} 
\e^{2\rho}= \frac{1}{a^{2} }\left( -\mu + a^{2} - \frac{a^4}{l_\mathrm{dS}^2} \right) \, , 
\end{equation} 
we obtain 
\begin{equation} 
\label{H4} H^2= \frac{1}{ l_\mathrm{dS}^2} - \frac{1}{a^2} + \frac{\mu}{a^4}
+ \frac{\kappa^2}{4} \, .
\end{equation}
Here $l_\mathrm{dS}$ is the curvature radius of the de Sitter space-time and $\mu$ is the black 
hole mass.
On the other hand, in the Schwarzschild-AdS black hole, 
\begin{equation} 
\label{AdSS} 
\e^{2\rho}= \frac{1}{a^{2} }\left( -\mu + a^{2} + \frac{a^4}{l_\mathrm{AdS}^2} \right) \, .
\end{equation}
we obtain,
\begin{align}
\label{F01}
H^2= - \frac{1}{l_\mathrm{AdS}^2} - \frac{1}{a^2}
+ \frac{\mu}{a^4} + \frac{\kappa^2}{4}\, .
\end{align}
In the Jordan frame, the metric is given by 
\begin{equation} 
\label{dsJ} 
ds^2_{\mathrm{J}4} = {F^{(5)}}'(R) ds^{2}_4=\left( -d\tau ^2 +a^2 d
\Omega_\mathrm{3}^2 \right)\, .
\end{equation}
Because the scalar curvature is a constant in the Schwarzschild-(anti-)de Sitter space-time, 
${F^{(5)}}'(R)$ can be absorbed into the redefinition of $\tau$ and $a$, 
\begin{equation} 
\label{dsJ2} 
d\tilde\tau \equiv dt \sqrt{{F^{(5)}}'(R)} \, , \quad \tilde a \equiv a \sqrt{{F^{(5)}}'(R)} \, .
\end{equation}
Then the qualitative properties are not changed in the Jordan frame compared with the Einstein 
frame.
We should also note that the motion of the brane does not depend on the detailed structure of 
$F^{(5)}(R)$.

In the Nariai space, the radius $a$ is a constant and therefore $H=0$.
Furthermore in the Nariai space, we find $\e^{2\rho(a)}=0$ and therefore
Eq.~(\ref{HH0}) shows that the brane tension $\tilde\kappa$ should vanish.
That is, if and only if the tension vanished, the brane can exist.
The non-vanishing tension might be cancelled with the contribution from the trace anomaly by 
tuning the brane tension.
We should note, however, that there should not be any (FRW) dynamics of the brane in the Nariai 
space.

However, the anti-evaporation may induce the dynamics of the brane.
For the metric (\ref{dse}), one gets the expressions of the connection in (\ref{Nconnections}).
We introduce a new time coordinate $\tilde t$ in the metric (\ref{dse}) as follows, 
\be 
\label{NFRW1} 
d\tilde t^2 \equiv \e^{2\rho} \left( d\tau^2 - dx^2 \right) \, .
\ee
Then the metric(\ref{dse}) reduces to the form of the FRW-like metric, 
\begin{equation} 
\label{NFRW2} 
ds^{2}= - d\tau^2 +\e^{-2\phi(x,\tau)}d\Omega_{(3)}^2 \, , 
\end{equation} 
if we identify $\e^{-\phi(x,\tau)}$ with the scale factor $a$, $a=\e^{-\phi(x,\tau)}$.
Then the unit vector perpendicular to the brane is given by 
\be 
\label{NFRW3} 
n^\mu =\left( - \e^{-2\rho} \frac{\partial x}{\partial \tilde t},
  -\e^{-2\rho} \frac{\partial \tau}{\partial \tilde t}, 0,0,0 \right)\, ,
\ee
and the $(i,j)$ $\left(i,j=1,2,3\right)$ components Eq.~(\ref{EE9}) give 
\be 
\label{NFRW4}
 - \e^{-2\rho} \frac{\partial \phi}{\partial \tau} \frac{\partial x}{\partial \tilde t}
 - \e^{-2\rho} \frac{\partial \phi}{\partial x} \frac{\partial \tau}{\partial \tilde t} = \tilde \kappa \, .
\ee
As we discussed, in order that the brane exists in the Nariai space-time, we find 
$\tilde \kappa = 0$.
By using the solution in (\ref{S1}), and analytically recontinuing the coordinates $x\to - i\tau$, 
$\tau \to - ix$, if we assume 
\be 
\label{NFRW5} 
\phi = \ln \Lambda + \phi_0 \cosh \omega \tau \cosh^\beta x \, , 
\ee 
with $\omega^2 = \beta^2$, we find 
\be 
\label{NFRW6}
 - \omega \sinh \omega \tau \cosh^\beta x \frac{\partial x}{\partial \tilde t}
 - \beta \cosh \omega \tau \cosh^{\beta-1} x \sinh x \frac{\partial \tau}{\partial \tilde t} = 0\, , 
 \ee 
 that is, 
\be 
\label{NFRW7} 
\frac{\partial x}{\partial \tilde t} 
= - \frac{\beta \tanh x}{\omega \tanh \omega \tau} \frac{\partial \tau}{\partial \tilde t} \, .
\ee
Assuming that $x$ and $\tau$ only depend on $\tilde t$ on the brane, 
\be 
\label{NFRW8}
0 = \frac{1}{\beta \tanh x}\frac{dx}{d\tilde t}
+ \frac{1}{\omega \tanh \omega\tau} \frac{d\tau}{d\tilde t}
= \frac{d}{d\tilde t} \left( \frac{1}{\beta} \ln \sinh x
+ \ln \tanh \omega \tau \right)\, ,
\ee
that is $\frac{1}{\omega} \ln \sinh x+ \ln \sinh \omega \tau$ is a constant, which gives the 
trajectory of  the brane, 
\be 
\label{NFRW9} 
\sinh x = \frac{C}{\sinh^\beta \omega \tau}\, .
\ee
Here $C$ is a constant.
Of course, the expression in (\ref{NFRW9}) is valid as long as the perturbation 
$\delta\phi = \phi_0 \cosh \omega \tau \cosh^\beta x$ is small enough.
We should also note that because ${F^{(5)}}'(R)$ is not a constant due to the perturbation,
Eq.~(\ref{dsJ2}) also gives another source of the dynamics of the brane. However, that 
Eq.~(\ref{dsJ2}) gives only small correction to Eq.~(\ref{NFRW9}).

\section{Discussions and open problems}

In this review we have discussed the evaporation and antievaporation phenomena within the framework of extended theories of gravity. In particular, we have indentified two particular metrics -- the Nariai and the extremal Reissner-Nordstr\"om black hole solutions --- that are unstable at the first order of metric perturbations. Explicit analyses with the cases of dilaton-gravity, $f(R)$-gravity, $f(T)$-gravity, Mimetic gravity and string-inspired gravity show up the emerging of the evaporation and antievaporation instabilities. We have seen how these instabilities change completely the thermodynamical behavior of black holes. The most surprising result is the suppression of the Bekenstein-Hawking radiation.\\

Several further questions may naturally arise. First of all, since (anti)evaporation instabilities seem to be so ubiquitous, we may ask whether any fundamental principle could be found, common to extended theories of gravity, that could motivate the emergence of such a phenomena. Second, since evaporation and antievaporation turn off Bekenstein-Hawking radiation, we may ask whether these phenomena can be relevant for the black hole information paradox.\\

Another unclear point remains the sensitivity of the evaporation/antievaporation transition on 
integration constants that seems undetermined by the initial conditions of the problem. Is there any principle to establish them? To use Hawking's words, is there a loss of predictability behind such a problem?

The last point is also crucially related to a possible cosmological problem. The production of primordial black holes, described by Nariai metrics, can lead to a disastrous cosmological instability. In fact, the antievaporation, turning off Bekenstein-Hawking emission, can lead to a catastrophic exponential expansion of primordial black holes. This is an issue that still needs to be better understood in the literature. 
We emphasize indeed that the implications of these phenomena on the information loss paradox and on the holographic principle have not yet discussed in litterature. In other words, the interpretation of such instabilities of the black hole in the bulk has not a clear interpretation on the boundary theory. \\

(Anti)evaporation may be related to a way-out from the 
the Firewall paradox \cite{Almheiri:2012rt,Braunstein:2009my}. 
The Firewall paradox is originated from the holographic entanglement 
among the black hole interior and the emitted Bekenstein-Hawking radiation \cite{Almheiri:2012rt,Braunstein:2009my}. 
This leads to a paradoxical violation of unitarity in quantum mechanics as 
well as the Equivalence Principle of General Relativity. 
However, the (anti)evaporation instability radically changes 
the black hole emission, leading to a suppression of the Bekenstein-Hawking 
radiation -- as mentioned above. So that, it is conceivable that 
the (anti)evaporation carries deep consequences in our understanding of the black hole 
information paradox. \\

On the other hand, it is still unknown if exotic black holes solutions
with multi-event horizons
 of alternative theories of gravity 
in presence of a non-linear electrodynamical field -- like the one recently found in Ref. \cite{Nojiri:2017kex} --
can have (anti-)evaporation in some regions of the parameter space. \\

 In the era of gravitational waves discovery from the LIGO collaboration
\cite{Abbott:2016blz,Abbott:2016nmj,Abbott:2017vtc},
crucial informations on the (anti-)evaporation phenomena 
can be provided from informations on the gravitational waves signals 
signal, searching for deviations from general relativity predictions \cite{Corda:2009re}.

We can then conclude that evaporation and antievaporation instabilities are interesting new phenomena that cannot be found in standard General Relativity, but are common in many of its extensions. Several deep issues could still be hidden behind them, and for the moment being it seems still we are still far from a complete understanding of their implications. 


\begin{acknowledgments}
\noindent 
A.A. thanks Salvatore Capozziello and Sergei Odintsov for many discussions on these aspects. A.M. wishes to acknowledge support by the Shanghai Municipality, through the grant No. KBH1512299, and by Fudan University, through the grant No. JJH1512105.

\end{acknowledgments}


\appendix

\section{Nariai metric \label{A1}}

The Nariai metric can be obtained as a particular
limit of the Schwarzschild-de Sitter (SdS) black hole.
Let's start from the SdS solution in four space-time dimensions:
\be\label{ds}
ds^{2}=-\Phi(R)dt^{2}+\frac{dr^{2}}{\Phi(r)}+r^{2}d\Omega^{2}
\ee
where $d\Omega^{2}=\sin\psi^{2}d\theta^{2}+d\psi^{2}$ 
and 
\be\label{Phi}
\Phi(r)=1-\frac{2M}{r}-\frac{\lambda}{3}r^{2}
\ee
where $M$ is the Black hole mass and $\lambda$ is a cosmological constant. 

Now, in order to smoothly perform the limit to the extremal SdS black hole, 
where $9M^{2}\lambda \rightarrow 1$, 
let us introduce two new coordinates $\psi',\chi$
as follows:
\be\label{tr}
t=\frac{1}{\epsilon \Lambda }\psi',\,\,\,r=\frac{1}{\Lambda}\left[1-\epsilon \cos \chi-\frac{1}{6\epsilon^{2}}\right]\, ,
\ee 
where $9M^{2}\lambda=1-3\epsilon^{2}$, 
$0 \leq\epsilon <<1$,
$\lambda=\sqrt{\Lambda}$.
The new metric in this coordinate system is 
\be\label{SdS}
ds^{2}=-\frac{1}{\Lambda^{2}}\left( 1+\frac{2}{3}\epsilon \cos \chi\right)\sin^{2}\chi d\psi'^{2}+\frac{1}{\Lambda^{2}}\left( 1-\frac{2}{3}\epsilon \cos \chi\right)d\chi^{2}+\frac{1}{\Lambda^{2}}(1-2\epsilon \cos\chi)d\Omega^{2}\, .
\ee
which, in the limit of $\epsilon \rightarrow 0$, smoothly converges 
to the so dubbed Nariai space-time
\be\label{dsss}
ds^{2}=\frac{1}{\Lambda^{2}}(-\sin^{2}\chi d\psi'^{2}+d\chi^{2})+\frac{1}{\Lambda^{2}}d\Omega^{2}\, . 
\ee
At this point, one can introduce a series of coordinate changes.
First of all, 
we introduce $\chi=-\arcsin z$, 
such that 
\be\label{dsss}
ds^{2}=-\frac{1}{\Lambda^{2}}(1-z^{2})d\psi'^{2}+\frac{dz^{2}}{\Lambda^{2}(1-z^{2})}+\frac{1}{\Lambda^{2}}d\Omega^{2}\, .
\ee
Then we can write the cosmological time variable and the comoving coordinate 
$x$ as
\be\label{tx}
t=\psi'+\frac{1}{2}\log (1-z^{2}),\,\,\,\, x=\frac{z}{(1-z)^{1/2}}e^{\pm t}
\ee
leading to the metric 
\be\label{gene}
ds^{2}=\frac{1}{\Lambda^{2}}(-dt^{2}+\cosh^{2}t dx^{2}+d\Omega^{2}),
\ee
as a linear combination of metrics 
\be\label{com}
ds^{2}=\frac{1}{\Lambda^{2}}(-dt^{2}+e^{\pm 2t}dx^{2})+\frac{1}{\Lambda^{2}}d\Omega^{2}\,.
\ee
Finally, let us remark that with an analytic continuation of the time-coordinate,
one can transform the 
dependences on the $\cosh^{2}t$ to $\cos^{2}t$, as done previously in the litterature.

\section{Extremal Reissner-N\"ordstrom metric  \label{A2}}

In this section we will review the extremal Reissner-Nordstr\"om metric.
Let us start
Reissner-Nordstr\"om metric:
\be\label{RNN}
ds^{2}=-\Phi(r)dt^{2}+\Phi(r)^{-1}dr^{2}+r^{2}d\Omega^{2}\, ,
\ee 
where 
\be\label{Phirrrr}
\Phi(r)=1-\frac{R_{0}r^{2}}{12}-\frac{M}{r}+\frac{Q}{r^{2}}\, . 
\ee
We can rewrite the mass and the charge in terms of the two radius 
\be\label{MQQ}
Q=r_{0}r_{1}\left( 1-\frac{R_{0}(r_{0}^{2}+r_{1}^{2}+r_{0}r_{1})}{12}\right)\, 
\ee
and 
\be\label{phiiii}
\Phi(r)=\left(1-\frac{r_{0}}{r}\right)\left(1-\frac{r_{1}}{r}\right)\left\{1-\frac{R_{0}[(r+r_{0})(r+r_{1})+r_{0}^{2}+r_{1}^{2}]}{12}    \right\}\,.
\ee
Now, we can smoothly perform the limit of $r_{0}\rightarrow r_{1}$,
i.e. extremal limit.
We can consider the following coordinate change, 
\be \label{rurz}
r_{1}=r_{0}+\epsilon,\,\,\,r=r_{0}+\frac{\epsilon}{2}(1+\tanh x)\, .
\ee
For $\epsilon \rightarrow 0$, 
\be\label{AAAAA}
\Phi \rightarrow -\frac{\epsilon^{2}}{4 r_{0}^{2}}\left( 1-\frac{r_{0}^{2}R_{0}}{2}\right)\cosh^{2x}\, . 
\ee
By redefining $t$ as
\be\label{ttt}
t=\frac{2r_{0}^{2}}{\epsilon \left(1-\frac{r_{0}^{2}R_{0}}{2} \right)}\tau \, , 
\ee
one obtains 
\be\label{Nasisiis}
ds^{2}=\frac{r_{0}^{2}}{\left(1-\frac{r_{0}^{2}R_{0}}{2} \right)\cosh^{2}x}(d\tau^{2}-dx^{2})+r_{0}^{2}d\Omega^{2}\, .
\ee

\section{Components of the Ricci tensors and Ricci scalar of 4D Nariai Black holes in $f(R)$-gravity  \label{A3}}

\be\label{GGG}
\Gamma_{\tau\tau}^{\tau}=\Gamma_{xx}^{\tau}=\Gamma_{\tau x}^{x}=\dot{\rho}\,,\,\,\,\,\Gamma_{xx}^{x}=\Gamma_{\tau \tau}^{x}=\Gamma_{x\tau}^{\tau}=\rho'\, ,
\ee
\be\label{GG}
\Gamma_{\psi\psi}^{\tau}=\Gamma_{\theta\theta}^{\tau}\sin^{2}\theta=-\dot{\phi}e^{-(\rho+\phi)}\sin^{2}\theta \, , \Gamma_{\psi\psi}^{x}=\Gamma_{\theta\theta}^{x}\sin^{2}\theta=\phi' e^{-(\rho+\phi)}\sin^{2}\theta\, ,
\ee
\be\label{G}
\Gamma_{\tau \theta}^{\theta}=\Gamma_{\theta \tau}^{\tau}=\Gamma_{\psi \tau}^{\psi}=-\dot{\phi},\,\,\,
\Gamma_{x\theta}^{\theta}=\Gamma_{\theta x}^{\theta}=\Gamma_{x\psi}^{\psi}=\Gamma_{\psi x}^{\psi}=-\phi',\,\,\,
\Gamma_{\psi\psi}^{\theta}=-\sin \theta \cos \theta,\,\,\,\Gamma_{\psi\theta}^{\psi}=\Gamma_{\theta\psi}^{\psi}=\cot \theta\,,
\ee
\be\label{RRRR}
R_{\tau\tau}=-\ddot{\rho}+2\ddot{\phi}+\rho''-2\dot{\phi}^{2}-2\dot{\rho}\dot{\phi}-2\rho'\phi',\,\,\,
R_{xx}=-\rho''+\ddot{\rho}+2\phi''-2\phi'^{2}-2\dot{\rho}\dot{\phi}-2\rho'\phi'\,,
\ee
\be\label{RRRRR}
R_{x\tau}=R_{\tau x}=2\dot{\phi}'-2\phi' \dot{\phi}-2\rho'\dot{\phi}-2\dot{\rho}\phi',\,\,\,
R_{\phi\phi}=R_{\theta \theta}\sin^{2}\theta=\left\{1+e^{-2(\rho+\phi)}\left(-\ddot{\phi}+\phi''+2\dot{\phi}^{2}-2\phi'^{2}\right)\right\}\sin^{2}\theta
\ee
\be\label{RRRRRR}
R=(2\ddot{\rho}-2\rho''-4\ddot{\phi}+4\phi''+6\dot{\phi}^{2}-6\phi'^{2})e^{-2\rho}+2e^{2\phi}\,.
\ee

\section{Components of the Ricci tensors and Ricci scalar in extremal Reissner-N\"ordstrom BH \label{A4}}

\be
\label{Gammaaa}
\Gamma_{\tau\tau}^{\tau}=\Gamma_{xx}^{\tau}=\Gamma_{\tau x}^{\tau}=\Gamma_{x\tau}^{\tau}=\dot{\rho},\,\,\,\,\Gamma_{xx}^{x}=\Gamma_{\tau\tau}^{x}=\Gamma_{\tau x}^{\tau}=\Gamma_{x\tau}^{\tau}=\rho' \, ,
\ee
\be
\label{Gammaxx}
\Gamma_{\psi\psi}^{\tau}=\Gamma_{\theta\theta}^{\tau}\sin^{2}\theta=-\frac{\Lambda^{2}}{\Lambda'^{2}}\dot{\phi}e^{-(\rho+\phi)}\sin^{2}\theta,\,\,\,\,\Gamma_{\phi\phi}^{x}=\Gamma_{\theta\theta}^{x}\sin^{2}\theta=\frac{\Lambda^{2}}{\Lambda'^{2}}\phi'e^{-(\rho+\phi)}\sin^{2}\theta\, ,
\ee
\be
\label{Gammathht}
\Gamma_{\tau \theta}^{\theta}=\Gamma_{\theta \tau}^{\theta}=\Gamma_{\tau \psi}^{\psi}=\Gamma_{\psi\tau}^{\psi}=-\dot{\phi},\,\,\,\, \Gamma_{x\theta}^{\theta}=\Gamma_{\theta x}^{\theta}=\Gamma_{x\psi}^{\psi}=\Gamma_{\psi x}^{\psi}=-\phi'\,, 
\ee
\be 
\label{Gammemm}
\Gamma_{\psi\psi}^{\theta}=-\sin \theta \cos \theta,\,\,\,\,\Gamma_{\psi \theta}^{\psi}=\Gamma_{\theta \psi}^{\psi}=\cot \theta\, ,
\ee

\be
\label{Rtt}
R_{\tau\tau}=-\ddot{\rho}+2\ddot{\phi}+\rho''-2\dot{\phi}^{2}-2\dot{\rho}\dot{\phi}-2\rho'\phi',\,\,\,\,R_{xx}=-\rho''+\ddot{\rho}+2\phi''-2\phi'^{2}-2\dot{\rho}\dot{\phi}-2\rho'\phi'\,, 
\ee
\be \label{RRR}
R_{\tau x}=R_{x\tau}=2\dot{\phi}'-2\phi'\dot{\phi}-2\rho'\dot{\phi}-2\dot{\rho}\phi',\,\,\,R_{\psi\psi}=R_{\theta\theta}\sin^{2}=\left\{ 
1+\frac{\Lambda^{2}}{\Lambda'^{2}}e^{-2(\rho+\phi)}(-\ddot{\phi}+\phi''+2\dot{\phi}^{2}-2\phi'^{2})\right\}\sin^{2}\theta\, ,
\ee
\be \label{RRR}
R=\Lambda^{2}(2\ddot{\rho}-2\rho''-4\ddot{\phi}+4\phi''+6\dot{\phi}^{2}-6\phi'^{2})e^{-2\rho}+\Lambda'^{2}e^{2\phi}\, .
\ee

\section{Components of the Ricci tensors and Ricci scalar \label{A5} in five-dimensional Nariai black holes}

For the metric (\ref{dse}),
we write $\eta_{ab}=\mathrm{diag}(-1,0)$, $\left(a,b=\tau,x\right)$.
For the metric $d\Omega_{(3)}^2$ of three dimensional unit sphere, we also write as 
\be 
\label{3dS}
d\Omega_{(3)}^2 = \hat g_{ij} dx^i dx^j \, , \quad (i,j=1,2,3)\, .
\ee
Then we obtain $\hat R_{ij} = 2 \hat g_{ij}$.
Here $\hat R_{ij}$ is the Ricci curvature given by $\hat g_{ij}$.

Then we find the following expression of the connections, 
\be 
\label{Nconnections} 
\Gamma^a_{bc} = \delta^a_{\ b} \rho_{,c} + \delta^a_{\ c} \rho_{,b}
 - \eta_{bc} \rho^{,a}\, , \quad
\Gamma^a_{ij} = e^{-2(\rho+\phi)} \hat g_{ij} \phi^{,a}\, , \quad 
\Gamma^i_{aj} = \Gamma^i_{ja} = - \delta^i_{\ j} \phi_{,a}\, , \quad 
\Gamma^i_{jk} = \hat \Gamma^i_{jk}\, .
\ee
Here $\hat \Gamma^i_{jk}$ is the connection given by $\hat g_{ij}$.
By using the expressions in (\ref{Nconnections}), the curvatures are given by 
\be\label{Ncurvature} 
R_{ab} =3 \phi_{,ab} - \eta_{ab} \partial^{2} \rho - 3 \left(\phi_{,a} \rho_{,b}
+ \phi_{,b} \rho_{,a} \right) + 3 \eta_{ab} \phi_{,c} \rho^{,c}
 - 3 \phi_{,a} \phi_{,b} \, , 
 \ee
$$R_{ij} = \hat R_{ij} + \hat g_{ij} e^{-2 (\rho+\phi)} \left( \partial^{2} \phi
 - 3 \phi_{,a} \phi^{,a} \right)\, , \quad R_{ia}=R_{ai}=0\, , $$
 $$R = e^{2\phi} \hat R
+ \e^{-2\rho} \left( 6 \partial^{2} \phi - 2 \partial^{2} \rho - 12 \phi_{,a} \phi^{,a} 
\right) \, .$$
We should note $\hat R=6$ because $\hat R_{ij} = 2 \hat g_{ij}$.

\end{document}